\documentclass[aps,pra,twocolumn,showpacs,superscriptaddress,groupedaddress]{revtex4}

\usepackage{graphicx}
\usepackage{amsthm}
\usepackage[toc,page,title,titletoc,header]{appendix}
\usepackage{graphicx}
\usepackage{epstopdf}
\usepackage{CJK}

\theoremstyle{plain}% default

\theoremstyle{definition}

\theoremstyle{remark}

\newcommand{\ket}[1]{|#1\rangle}
\newcommand{\bra}[1]{\langle#1|}
\begin{document}

\begin{CJK*}{GB}{gbsn}

\title{Adiabatic evolution of decoherence-free subspaces and its shortcuts}

\author{S. L. Wu(ÎäËÉÁÖ)}
\email{slwu@dlnu.edu.cn}
\affiliation{School of Physics and Materials Engineering,
Dalian Nationalities University, Dalian 116600 China}

\author{X. L. Huang (»ÆÏþÀí)}
\affiliation{School of Physics and Electronic Technology,
Liaoning Normal University, Dalian 116029, China}

\author{H. Li(Àîºê)}
\affiliation{Center for Quantum Sciences and School of Physics,
Northeast Normal University, Changchun 130024, China}

\author{X. X. Yi(ÒÂѧϲ)}
\email{yixx@nenu.edu.cn}
\affiliation{Center for Quantum Sciences and School of Physics,
Northeast Normal University, Changchun 130024, China}

\date{\today}

\begin{abstract}
The adiabatic theorem and shortcuts-to-adiabaticity
for  time-dependent open quantum systems are explored in this paper. Starting from the definition of dynamical stable decoherence-free subspace, we show
that, under a compact adiabatic condition, the quantum state remains  in
the time-dependent decoherence-free subspace
with an extremely high purity, even though the dynamics of the open quantum system may be not adiabatic. The adiabatic condition mentioned here in the adiabatic theorem for open systems is very similar with that for closed quantum systems, except that the  operators required to change slowly are the Lindblad operators. We also show that the adiabatic evolution of decoherence-free subspaces (\emph{will be refereed to  as adiabatic decoherence-free subspaces(ADFSs) later}) depends on the existence of instantaneous decoherence-free subspaces,
which requires that the Hamiltonian  of open quantum systems has to be
engineered according to the incoherent control protocol. Besides,
the shortcuts-to-adiabaticity for the  adiabatic decoherence-free subspaces
is also presented based on the transitionless quantum driving method.
Finally, we provide an example that consists of  a two-level system  coupled to a
broad band squeezed vacuum field  to show our theory.
Our approach employs Markovian master equations and the theory can apply
to finite-dimensional quantum open systems.
\end{abstract}

 \maketitle

\end{CJK*}

\section{Introduction}

The adiabatic theorem is among the most useful results that has been known since the early days of quantum mechanics \cite{adi1-1,adi1-2}. The theorem states that if a quantum system is in an instantaneous eigenstate of  its time-dependent  Hamiltonian $H(t)$ at one time,  it will remain  to that eigenstate up to a phase factor at later times, provided that its Hamiltonian changes very slowly.  Recent developments in quantum information processing makes   the quantum adiabatic dynamics active again, since the adiabativity  possesses   intrinsic robustness gainst  the control errors in the Hamiltonian \cite{adicom1,adicom2}. However, for a long time the adiabatic theorem is almost exclusively concerned with closed systems. To take the  couplings of  qubits to environments into account, many efforts have  been recently put forward  to extending  the adiabatic theorem from close systems to open systems.  For example,  to formulate the adiabatic theorem for open quantum systems by the Jordan block decomposition of the dissipative generator \cite{adio1,adio1-1}, in the weak coupling limit\cite{adio2,adio3}, at zero temperature \cite{adio4},  by the method of effective Hamiltonian \cite{adio4-1}, by the noiseless subsystem decomposition \cite{adio5}, or in terms of  the instantaneous steady state of the Liouvillian\cite{adi2}.

Among those enlightening approaches, there is an approach that combines the adiabatic dynamics with decoherence-free subspaces, known as adiabatic evolution of decoherence-free subspaces (ADFSs)\cite{adfs1}. The existence of an ADFS is by no means trivial, and its presence often reflects a symmetry preserving evolution. When the symmetry preserving evolution is slow enough, the state of open quantum systems lying initially  in a time-dependent decoherence-free subspace(t-DFS) remains inside the subspace in latter times, and it is rigidly transported in the Hilbert space together with the t-DFS. However,  there is no answer to the question: \emph{How slow the  evolution can guarantee the quantum state inside  the t-DFS with extremely high purity.} In other words, the adiabatic condition for t-DFS is still missing, although the t-DFS was generalized into nonadiabatic case\cite{adfs2,tdfs1}. As shown, the coherent control is significant to stabilize quantum states in the t-DFS.  Therefore, we may ask the other question: \emph{ whether the ADFSs  also needs the assistance of coherent control? }

In this paper, we will explore the adiabatic theorem for t-DFSs  and present a suitable adiabatic condition to answer those questions. We have the following observations. Firstly, the adiabatic theorem shows that the coherent control on  open quantum systems is crucial for the ADFSs.   Both the incoherent  and coherent evolutions  are necessary for  the quantum state to stay  steadily in the t-DFSs.   Secondly, the adiabatic condition, which has similar form with the adiabatic condition for closed quantum systems\cite{adi1-2}, reveals  that
the quantum state would keep in the t-DFSs with extremely high purity, if the change rate of the t-DFSs is very small with respect to  the effective frequency difference of the non-hermitian Hamiltonian.
Even if the effective Hamiltonian of the open quantum system  is degenerate, the
ADFSs  can still be realized due to the decoherence, and the condition is available in this case as well.

Following the steps of the adiabatic theorem of closed quantum systems, we will derive adiabatic conditions on operators of the open quantum system. Since the t-DFSs are spanned by a set of common degenerate
eigenstates of Lindblad operators, the motion of t-DFSs attributes to the  symmetry preserving evolution of the environment. This can be realized by the incoherent control protocol leading to  the adiabatic condition on the
Lindblad operators. The adiabatic condition on the Lindblad operators shows that  the Lindblad operators have to change so slowly that the quantum state follows the t-DFS without  purity loss.

Recall that an instantaneous eigenstate  of a time-dependent Hamiltonian can evolve to the other instantaneous eigenstate at later time, \textbf{which is faster than the adiabatic evolution by shortcuts-to-adiabaticity (STA) in closed systems\cite{chen,chen-r}}. Besides,  it can also be realized by the other \textbf{particular techniques within the
broad concept of shortcuts to adiabaticity} such as the transitionless quantum driving method\cite{berry,Demirplak1,Demirplak2,Demirplak3, nonmx}, the
inverse engineering program\cite{chen}, and the fast quench dynamics\cite{fq}. It is then interesting to ask  \emph{whether the shortcuts-to-adiabaticity can be extended into   open quantum systems? }As far as we know, some works has been devoted on this topic, such as the transitionless quantum driving method\cite{opens} and the inverse
engineering program\cite{nonm}. In this paper, we will  consider how to accelerate the evolution of a state in the t-DFSs, and  by using Berry's transitionless quantum driving method, we will propose a protocol for the shortcuts-to-adiabaticity  open systems.  The results show that  the ADFSs  can be accelerated
by modulating  the coherent evolution\cite{tdfs1}.

The remainder of this paper is organized as follows. In Sec. \ref{sub:adfs},  begin with the definition of the dynamical stable DFSs, we present an adiabatic theorem and an   condition for the ADFSs. In Sec.\ref{sub:adic1}, we apply the adiabatic condition into the Lindblad operators and derive a condition on these operators. Instead of examining whether  the  condition for ADFSs is sufficient or necessary, in Sec.\ref{sub:The-Lower-Bound} we present a lower bound for the purity to quantify the system in the t-DFS. Based on the transitionless quantum driving method, we propose a program to accelerate the adiabatic dynamics of t-DFSs in Sec.\ref{shortcuts}. To illustrate  the theory, we present  a concrete example in Sec. \ref{sec3}. The effect of the coherent evolution and the decay  on the adiabatic dynamics is also numerically explored in this section. Finally, we conclude  in Sec. \ref{conclusion}.

\section{Adiabatic evolution of t-DFS}

\subsection{A theorem for ADFSs }\label{sub:adfs}

In this section, we will present a sufficient and necessary condition for dynamical
stable decoherence free subspaces\cite{dsdfs}. Let us consider
an open quantum system of  $N$-dimension  described by the following
master equation,
\begin{eqnarray}
\partial_{t}\hat{\rho}(t) & = & \hat{\mathcal{L}}(\hat{\rho}),\nonumber \\
\hat{\mathcal{L}}(\hat{\rho}) & = & -i[\hat{H}_0,\hat{\rho}]+\sum_{\alpha}
\left(\hat{F}_{\alpha}\hat{\rho}\hat{F}_{\alpha}^{\dagger}
-\frac{1}{2}\{\hat{F}_{\alpha}^{\dagger}\hat{F}_{\alpha},
\hat{\rho}\}\right).\label{eq:maeq}
\end{eqnarray}
where $\hat H_0(t)$ and $\{\hat F_\alpha (t)\}$ are the Hamiltonian and Lindblad operators in \textbf{an interaction picture.}
Based on this master equation, a dynamical stable DFS $\mathcal{H}_{\text{DFS}}$
is defined as a collection of quantum states such that the pure state
$\rho(t)\in\mathcal{H_{\text{DFS}}}$ fulfills
\begin{eqnarray}
\frac{\partial}{\partial t }p(t)=\frac{\partial}{\partial t }\text{Tr}[\rho^{2}(t)] =  0,\label{eq:purity}
\end{eqnarray}
which leads to the following conditions for DFSs :
The space
\[
\mathcal{H}_{\text{DFS}}:=\text{span}\{|\Phi_{1}\rangle,|\Phi_{2}\rangle,...,|\Phi_{M}\rangle\}
\]
is a $M$-dimensional DFS if and only if the bases fulfill following conditions: (1) The orthogonal bases of the DFS are the degenerate eigenstates of the Lindblad operators, i.e.,
\begin{equation}
\hat{F}_{\alpha}|\Phi_{j}\rangle=c_{\alpha}|\Phi_{j}\rangle,\ \forall\alpha,\ j;\label{a}
\end{equation}
 (2) $\mathcal{H}_{\text{DFS}}$ is invariant under the effective Hamiltonian
\begin{equation}
\hat{H}_{\text{eff}}^{0}=\hat{H}_0+\frac{i}{2}\sum_{\alpha}
\left(c_{\alpha}^{*}\hat{F}_{\alpha}-c_{\alpha}\hat{F}_{\alpha}^{\dagger}\right),\label{heff}
\end{equation}
which means that the quantum state in the DFS is still a state in such DFS
after the acting  of $\hat{H}_{\text{eff}}^0$, i.e.,
\begin{equation}
\langle\Phi_{n}^{\bot}|\hat{H}_{\text{eff}}^{0}|\Phi_{j}\rangle=0,\,\forall n,\,j,\label{eq:heff}
\end{equation}
where $|\Phi_{n}^{\bot}\rangle$ is the $n$-th basis of complementary
subspace $\mathcal{H}_{\text{CS}}$.

{Here, we notice that the bases of the DFS are common degenerate
eigenstates of Lindblad operators. Notice that  the Lindblad operators are not
always hermitian,  the eigenstates of the Lindblad operators may be
nonorthogonal. Thus, the biorthogonal bases $\{\left|\Phi_i\right),
\left|\Psi_i\right)\}$ have to be used\cite{bior}. Assume that  $\left|\Phi_i\right)$
($\left|\Psi_i\right)$) is the $i$-th common degenerate right (left) eigenstate
of Lindblad operators with same eigenvalue $c_\alpha$,
the right eigenstates are nonorthogonal with each other, but are orthogonal to
the left eigenstates, i.e., $\left(\Psi_i|\Phi_j\right)=\delta_{ij}$.
By  the Schmidt orthogonalization, we could orthogonalize
the right eigenstates with the common eigenvalue $c_\alpha$, which is
the states $\{|\Phi_i\rangle\}$ used as the bases of DFSs in Eqs.(\ref{a})
and (\ref{eq:heff}).  It is obvious that the  biorthogonal bases set $\{\left|\Phi_i\right),
\left|\Psi_i\right)\}$ is equivalent to the  orthogonal set $\{|\Phi_i\rangle\}$,
because either $\left|\Phi_i\right)$ or $|\Phi_i\rangle$ is the common degenerate
eigenstates of $\hat F_\alpha$. Therefore, we use the orthogonal degenerate eigenvalues set
$\{|\Phi_i\rangle\}$ as the bases of DFSs in the following discussion.
On the other hand, even though the right eigenstates of Lindblad
operators are not orthogonal with each other, the set of all right
eigenstates is still a complete set of the total Hilbert space. Thus,
the rest of right eigenstates of Lindblad operators can be used, by  the
Schmidt orthogonalization again, to
determine bases of the complementary subspaces.}

In the following, we consider that Lindblad operators
presented in Eq.(\ref{eq:maeq}) are time-dependent, which might come from
the engineering parameters in the environment
(e.g., incoherent control scheme)\cite{ree} or in the  environment-system couplings\cite{nonm}. If there is a instantaneous subspace
$\mathcal{H}_{\text{DFS}}(t):=\text{span}\{|\Phi_{1}(t)\rangle,|\Phi_{2}(t)\rangle,...,|\Phi_{M}(t)\rangle\}$
which fulfills
\begin{equation}
\hat{F}_{\alpha}(t)|\Phi_{j}(t)\rangle=c_{\alpha}(t)|\Phi_{j}(t)\rangle,\label{eq:Falpha}
\end{equation}
and
\begin{equation}
\langle\Phi_{n}^{\bot}(t)|\hat{H}_{\text{eff}}(t)|\Phi_{j}(t)\rangle=0,\label{eq:Heff}
\end{equation}
the subspace $\mathcal{H}_{\text{DFS}}(t)$ is known as a time-dependent DFS(t-DFS) of the open quantum system. It has been shown that, when the eigenvalues
$c_{\alpha}(t)$ equal to zero, a t-DFS can be used to formulate  ADFSs without
further condition except the adiabatic one\cite{adfs1,adfs2}. In the following, we generalize the ADFSs  into the case where the eigenvalues $c_\alpha(t)$ are not zero and time-dependent.

We are now ready to informally state the  theorem for ADFSs:
\emph{If a pure quantum state is initialized at $t=0$ in $\mathcal{H}_{\text{DFS}}(0)$,
the final state at $t=T$ will be
in $\mathcal{H}_{\text{DFS}}(T)$, provided the bases of the t-DFS and
its complementary subspace fulfill }
\begin{equation}
\Xi(t):=\text{Max}_{\{n,i\}}\left|\frac{4\langle\Phi_{n}^{\bot}(t)|\partial_{t}|\Phi_{i}(t)\rangle}
{\omega_{ni}+i\Gamma_n}\right|\ll1,\,\forall n,i,\label{eq:maxnj}
\end{equation}
\emph{where $\omega_{ni}=\langle\Phi_n^\bot(t) |\hat{H}_{\text{eff}}^{0}(t)|\Phi_n^\bot(t)\rangle-
\langle\Phi_i(t) |\hat{H}_{\text{eff}}^{0}(t)|\Phi_i(t) \rangle$ and $\Gamma_n=\sum_\alpha
\langle\Phi_n^\bot(t) |\tilde F^\dag_\alpha(t)\tilde F_\alpha(t)|\Phi_n^\bot(t)\rangle$/2} with
$\tilde F_\alpha(t)=\hat F_\alpha(t)-c_\alpha(t)$.
The details to derive  the adiabatic condition in Eq.(\ref{eq:maxnj})
can be found in Appendix \ref{sec:Appendix-A}. \textbf{The adiabatic condition obtained here is similar
to the condition for  the adiabatic dynamics with the non-Hermitian Hamiltonian\cite{nonh}.}

We should emphasize that the  theorem for ADFSs obtained here is valid if there is a t-DFS $\mathcal{H}_{\text{DFS}}(t)$ at arbitrary
moment and the bases of the  t-DFS are continuous with time. Different from
the configuration of Ref.\cite{adfs1}, the instantaneous subspace
does not only require that the quantum state in $\mathcal{H}_{\text{DFS}}(t)$
is an eigenstate of Lindblad operators (see Eq.(\ref{eq:Falpha})),
but also has to satisfy the condition that  the quantum state in the t-DFS is still a state in t-DFS
after the acting of $\hat{H}_{\text{eff}}^0$ (see Eq.(\ref{eq:Heff})). Plugging  Eq.(\ref{eq:heff})
into Eq.(\ref{eq:Heff}), we obtain
\begin{equation}
\langle\Phi_{j}(t)|\hat{H}_0(t)|\Phi_{n}^{\bot}(t)\rangle=
-\frac{i}{2}\sum_{\alpha}c_{\alpha}(t)\langle\Phi_{j}(t)|
\hat{F}_{\alpha}(t)|\Phi_{n}^{\bot}(t)\rangle,\label{eq:HF}
\end{equation}
which is the very relationship between the coherent evolution  and the
incoherent evolution. This restriction is very important in realizing
ADFSs.

As shown in Ref.\cite{dsdfs}, the quantum state $|\varphi(t)\rangle$
in $\text{\ensuremath{\mathcal{H}}}_{\text{DFS}}(t)$ fulfills
\begin{equation}
\frac{\partial}{\partial t}|\varphi(t)\rangle=-i\hat{H}_{\text{eff}}^{0}(t)|\varphi(t)\rangle.\label{shor}
\end{equation}
The operator $\hat{H}_{\text{eff}}^{0}(t)$ plays a role of  effective
Hamiltonian for the open quantum system, when quantum states are  in
the t-DFSs. To this extent and  from the viewpoint of quantum control, a coherent
control satisfying Eq.(\ref{eq:HF}) can enable   the quantum state
to stay  in the t-DFSs.

The adiabatic condition presented in Eq.(\ref{eq:maxnj}) is very similar with the adiabatic condition for closed quantum systems\cite{adi1-1, adi1-2}, except that the decoherence  has been considered in the adiabatic condition. It has been verified that  the quantitative condition is insufficient in guaranteeing the validity of the adiabatic approximation for closed systems. In fact, the adiabatic condition for ADFSs does also face this difficulty. Therefore, in Sec. \ref{sub:The-Lower-Bound}, we will present the lower bound of purity to describe the performance of the adiabatic approximation, and show that the adiabatic condition is available, if the total evolution time is very large.

\subsection{Adiabatic condition on Lindblad operators}\label{sub:adic1}

Stimulating by  the adiabatic theorem for closed quantum systems, we may wonder whether
the adiabatic condition for open system
can be expressed in terms of   operators of the open quantum system.

Remember that $\{|\Phi_{i}(t)\rangle\}|_{i=1}^M$
are common degenerate eigenstates of $\hat{F}_{\alpha}(t)$ as defined in Eq.(\ref{a}).
A useful expression can be found by taking the time derivative of  Eq.(\ref{a}) and multiplying
the resulting expression by $\langle\Phi_{n}^{\bot}(t)|$,
\begin{eqnarray}
&&c_{\alpha}(t)\langle\Phi_{n}^{\bot}(t)|\partial_{t}|\Phi_{i}(t)\rangle\label{eq:cond1}\\
&&=\langle\Phi_{n}^{\bot}(t)|\partial_{t}\hat{F}_{\alpha}(t)|\Phi_{i}(t)\rangle+
 \langle\Phi_{n}^{\bot}(t)|\hat{F}_{\alpha}(t)\partial_{t}|\Phi_{i}(t)\rangle.\nonumber
\end{eqnarray}
Plugging
\begin{eqnarray}
\sum_{j=1}^M|\Phi_j(t)\rangle\langle\Phi_j(t)|+\sum_{m=1}^{N-M}
|\Phi_m^\bot(t)\rangle\langle\Phi_m^\bot(t)|=\hat I\nonumber
\end{eqnarray}
into the last term of Eq. (\ref{eq:cond1}), we obtain
\begin{eqnarray}
\langle\Phi_{n}^{\bot}(t)|\hat{F}_{\alpha}(t)&&\partial_{t}|\Phi_{i}(t)\rangle=
-\langle\Phi_{n}^{\bot}(t)|\hat{F}_{\alpha}(t)\hat{K}(t)|\Phi_{i}(t)\rangle \nonumber\\ &&+\langle\hat{F}_{\alpha}(t)\rangle_{n}\langle\Phi_{n}^{\bot}(t)|\partial_{t}|\Phi_{i}(t)\rangle,
\end{eqnarray}
with $\hat{K}=\sum_{m\neq n}|\Phi_{m}^{\bot}(t)\rangle\partial_{t}\langle\Phi_{m}^{\bot}(t)|$
and $\langle\hat{F}_{\alpha}(t)\rangle_{n}=\langle\Phi_{n}^{\bot}(t)|\hat F_{\alpha}(t)|\Phi_{n}^{\bot}(t)\rangle$.
Substituting the above expression back into
Eq.(\ref{eq:cond1}) and following the process in Appendix
\ref{appendix b}, we obtain
\begin{eqnarray}
\text{Max}_{\{n,i\}}&&\left|\frac{\langle\Phi_{n}^{\bot}(t)|\partial_{t}|\Phi_{i}(t)\rangle}
{\omega_{ni}+i\Gamma_n}\right|\leq
\nonumber\\&&\left(\frac{\sum_{a=0}^{N-M-1}P_{N-M}^{a+1}(F_{\text{Max}})^{a}}
 {N-M}\right)\nonumber\\&&\times\text{Max}_{\{n,i\}}
 \left|\frac{\langle\Phi_{n}^{\bot}|\partial_{t}\hat{F}_{\alpha}|\Phi_{i}\rangle}
 {(\omega_{ni}+i\Gamma_n)(c_{\alpha}-\langle\hat{F}_{\alpha}\rangle_{n})}\right|
 \nonumber\\\label{eq:adicond}
\end{eqnarray}
with a bounded real number
\begin{eqnarray*}
F_{\text{Max}}=\text{Max }_{\{n,m\}}\left|\frac{\langle\Phi_{m}(t)|\hat{F}_{\alpha}|\Phi_{n}^{\bot}(t)\rangle}
{c_{\alpha}-\langle\hat{F}_{\alpha}\rangle_{n}}\right|,
\end{eqnarray*}
and the number of permutation is $P_{N-M}^{a+1}$. To obtain Eq.(\ref{eq:adicond}),
we have assume that $c_{\alpha}\neq\langle\hat{F}_{\alpha}\rangle_{n}$
always holds.  The latter claim can be shown by rewriting the adiabatic condition as
\begin{eqnarray}
\Xi(t)&:=&
\text{Max}_{\{n,i\}}\left|\frac{4\langle\Phi_{n}^{\bot}|
\partial_{t}\hat{F}_{\alpha}|\Phi_{i}\rangle}{(c_{\alpha}-\langle\hat{F}_{\alpha}\rangle_{n})
(\omega_{ni}+i\Gamma_n)}
\right|\nonumber\\
&\times&\left(\frac{\sum_{a=0}^{N-M-1}P_{N-M}^{a+1}(F_{\text{Max}})^{a}}{N-M}\right)\ll1,\:\forall\alpha. \label{eq:c2}
\end{eqnarray}
with numbers of permutation $P_{N-M}^{a+1}$.

Generally speaking, the adiabatic theorem for open systems mentioned above can be expressed
as two independent conditions, i.e., the existence of the
t-DFSs and the adiabatic requirement on the motion of t-DFSs. The first
condition (\emph{the t-DFSs condition}) states that,
the t-DFSs are spanned by common degenerate eigenstates of the Lindblad operators(see Eq.(\ref{eq:Falpha})); and the Hamiltonian of open quantum systems should be engineered in order to keep the  quantum states in the t-DFSs (see Eq.(\ref{eq:HF})). This condition ensures
that the evolution of quantum states in the t-DFSs is unitary. Besides, this condition also claims that the requirement of adiabaticity on the Lindblad operators partly comes from the requirement on  the Hamiltonian. This can be found from Eq.(\ref{eq:HF}), it shows that when the change of Lindblad operators is adiabatic,  parts of $\hat H_0(t)$ have to change slowly to ensure that the t-DFSs condition holds. From the other viewpoint, the adiabatic requirement mentioned here is not only on the incoherent evolution(governed by Lindbladian), but also on the coherent evolution(governed by Hamiltonian).

The second condition (\emph{the adiabatic condition}) requires that the Lindblad operators have to change so slowly that the quantum state follows the t-DFS with extremely high purity(see Eq.(\ref{eq:c2})). In other words, the transition between the t-DFS and the complementary subspaces can be negligible, when the adiabatic condition Eq.(\ref{eq:c2}) is satisfied.
Moreover, there is the other term in the adiabatic condition, i.e. $\sum_{a=0}^{N-M-1}P_{N-M}^{a+1}(F_{\text{Max}})^{a}$. If this
term is finite, it does not affect the adiabatic condition
evidently. On the one hand, we may notice that, this term is
related to the size of complementary subspaces. The larger the complementary
subspace is, the more difficult the adiabatic condition to be held. On
the other hand, $F_{\text{Max}}$ describes the maximal transition induced
by Lindblad operators in the complementary subspace. If the Lindblad operators
$\{\hat{F}_{\alpha}\}$ do not induce any transition between the bases
of the complementary subspace ($F_{\text{Max}}=0$), the adiabatic condition
can be rewritten into a compact form (also see Eq.(\ref{eq:maxf})),
\begin{eqnarray}
\Xi(t)=
\text{Max}_{\{n,i\}}\left|\frac{4\langle\Phi_{n}^{\bot}
|\partial_{t}\hat{F}_{\alpha}|\Phi_{i}\rangle}
{(\omega_{ni}+i\Gamma_n)(c_{\alpha}-\langle\hat{F}_{\alpha}\rangle_{n})
}\right|\ll1\label{cond2}
\end{eqnarray}
In this case, the adiabatic condition  does not depend on the size
of complementary subspaces.

\subsection{Lower Bound of the Purity}\label{sub:The-Lower-Bound}

For closed quantum systems, the practical applications of the adiabatic
theorem rely on the  slowness of the Hamiltonian, which is also
required for the  theorem for ADFSs. Here, we  present
a lower bound for the purity to keep the system in the t-DFS, which is  a function of the total evolution
time $T$. As expected, the ADFSs  can be reached, if the total evolution time is long enough.

Firstly, we show that the t-DFSs can not be expanded by eigenstates
of the Hamiltonian $\hat{H}(t)$ of the open quantum system. Suppose
that part of eigenstates of $\hat{H}(t)$ can be used as bases of
the t-DFSs, i.e.,
\begin{eqnarray*}
\hat{H}_0(t)|\Phi_{j}(t)\rangle & = & \varepsilon_{j}(t)|\Phi_{j}(t)\rangle\\
\hat{H}_0(t)|\Phi_{n}^{\bot}(t)\rangle & = & \varepsilon_{n}^{\bot}(t)|\Phi_{n}^{\bot}(t)\rangle.
\end{eqnarray*}
Now we consider the condition for the ADFSs Eq. (\ref{eq:Heff}), it follows that,
\[
\langle\Phi_{n}^{\bot}|\hat{H}_{\text{eff}}^{0}(t)|\Phi_{j}(t)\rangle=
\sum_{\alpha}c_{\alpha}(t)\langle\Phi_{n}^{\bot}|\hat{F}^\dag_{\alpha}(t)|\Phi_{j}(t)\rangle\neq0
\]
As mentioned in Sec. \ref{sub:adfs}, the t-DFSs condition requires that the t-DFS is invariant under the operation of $\hat{H}_{\text{eff}}^{0}(t)$ (see Eq.(\ref{eq:Heff})). Therefore, the t-DFS can not be expanded by the eigenstate of $\hat{H}(t)$.

In fact, the eigenstates of $\hat{H}_{\text{eff}}^{0}(t)$ can be chosen
as the bases of the t-DFSs and its complementary subspaces, i.e.,
\begin{eqnarray*}
\hat{H}_{\text{eff}}^{0}(t)|\Phi_{i}(t)\rangle & = & E_{i}(t)|\Phi_{i}(t)\rangle,\\
\hat{H}_{\text{eff}}^{0}(t)|\Phi_{n}^{\bot}(t)\rangle & = & E_{n}^{\bot}(t)|\Phi_{n}^{\bot}(t)\rangle,
\end{eqnarray*}
where $E_{i}(t)$ ($E_{n}^{\bot}(t)$) is the eigenvalue corresponding
to the eigenstate $|\Phi_{i}(t)\rangle$ ($|\Phi_{n}^{\bot}(t)\rangle$).
It is easy to find that $\langle\Phi_{n}^{\bot}|\hat{H}_{\text{eff}}^{0}(t)|\Phi_{j}(t)\rangle=0$.
Hence the eigenstates of $\hat{H}_{\text{eff}}^{0}(t)$ can be divided into two bands: the ADFSs
bands and the complementary subspaces bands.

We start with the time derivative  of the purity Eq.(\ref{eq:tpur}).
It is convenient to choose eigenstates of the effective Hamiltonian
$\hat{H}_{\text{eff}}^{0}$ as the bases of  t-DFSs and its complementary
subspaces. Under adiabatic approximation, we rewrite Eq.(\ref{eq:tpur})
into scalar form
\begin{eqnarray}
\partial_{t}p(t) & \doteq & 2\text{Tr}\left\{\bar{\rho}_{D}\left(-i[\bar{G},\bar{\rho}_{N}]+\sum_{\alpha}
\tilde{F}_{\alpha}\bar{\rho}_{C}\tilde{F}_{\alpha}^{\dagger}\right)\right\},\nonumber \\
 & = & 4\text{Re}\left\{\sum_{ijm}\bar{\rho}_{ji}\bar{\rho}_{im}
 \langle\Phi_{m}^{\bot}|\partial_{t}|\Phi_{j}\rangle\right.\nonumber\\
 &  &\times\left.\exp\left(i\int_{0}^{t}(\omega_{mj}-i\Gamma_m)d\tau\right)\right\}\nonumber \\
 &  & +2\sum_{\alpha}\text{Tr}\left\{\bar{\rho}_{D}\tilde{F}_{\alpha}
 \bar{\rho}_{C}\tilde{F}_{\alpha}^{\dagger}\right\},\label{eq:pt}
\end{eqnarray}
where $\omega_{mj}=E^\bot_{m}-E_{j}$ and $\Gamma_m=\langle\hat\Gamma\rangle_m$. Following the same discussion
in Appendix \ref{sec:Appendix-A}, we discard the second term in Eq.(\ref{eq:pt}), and turn it
into an inequality,
\begin{eqnarray*}
\partial_{t}p(t)\geq&4&\text{Re}\left\{\sum_{ijm}\bar{\rho}_{ji}\bar{\rho}_{im}\langle\Phi_{m}^{\bot}|
\partial_{t}|\Phi_{j}\rangle\right.\nonumber\\
 &  &\times\left.\exp\left(i\int_{0}^{t}(\omega_{mj}-i\Gamma_m)d\tau\right)\right\},
\end{eqnarray*}
One may wonder  why the second term in Eq.(\ref{eq:pt}) can
be discarded. As discussed in Appendix \ref{sec:Appendix-A}, this term
describes the transition from the complementary subspaces into  the t-DFSs.
Since the population in complementary subspaces is very small under the
adiabatic approximation, the contribution from this term is negligible.
On the other hand, when the population leaks into the complementary
subspace, this term will push the quantum state back  to the t-DFS.
Therefore, this term will help to keep (even to increase) the system in the t-DFS, which can be understood as continuous  measurements caused zero effect
projecting the system onto the subspace\cite{zeno}. Therefore, if we discard
this term, the purity must be larger than the lower bound we present
here.

Recall  that $p(t)$ is given by
\begin{eqnarray*}
 p(t)-1\geq  &4&\text{Re}\left\{\sum_{ijm}\int_{0}^{t}\bar{\rho}_{ji}\bar{\rho}_{im}\langle\Phi_{m}^{\bot}|
 \partial_{t'}|\Phi_{j}\rangle\right.\nonumber\\
 &&\left.\times\exp\left(i\int_{0}^{t'}(\omega_{mj}-i\Gamma_m)d\tau\right)dt'\right\}.
\end{eqnarray*}
Integrating by part, we have
\begin{eqnarray}
p(t)-1&\geq&4\text{Re}\left\{\sum_{ijm}-i\bar{\rho}_{ji}\bar{\rho}_{im}\frac{\langle\Phi_{m}^{\bot}|\partial_{t}
 |\Phi_{j}\rangle}{\omega_{mj}+i\Gamma_{m}}\right.\nonumber\\
 &&\times\exp\left(i\int_{0}^{t}(\omega_{mj}+i\Gamma_{m})d\tau\right)\nonumber\\
 &&+i\int_{0}^{t}\partial_{t'}\bar{\rho}_{ji}\bar{\rho}_{im}\frac{\langle\Phi_{m}^{\bot}|\partial_{t'}
 |\Phi_{j}\rangle}{\omega_{mj}+i\Gamma_{m}}\nonumber\\
 &&\times\exp\left(i\int_{0}^{t'}(\omega_{mj}+i\Gamma_{m})d\tau\right)dt'\nonumber\\
 &&+i\int_{0}^{t}\bar{\rho}_{ji}\partial_{t'}\bar{\rho}_{im}\frac{\langle\Phi_{m}^{\bot}|\partial_{t'}
 |\Phi_{j}\rangle}{\omega_{mj}+i\Gamma_{m}}\nonumber\\
 &&\times\exp\left(i\int_{0}^{t'}(\omega_{mj}+i\Gamma_{m})d\tau\right)dt'\nonumber\\
 &&+i\int_{0}^{t}\bar{\rho}_{ji}\bar{\rho}_{im}\partial_{t'}(\frac{\langle\Phi_{m}^{\bot}|\partial_{t'}
 |\Phi_{j}\rangle}{\omega_{mj}+i\Gamma_{m}})\nonumber\\
 &&\left.\times\exp\left(i\int_{0}^{t'}(\omega_{mj}+i\Gamma_{m})d\tau\right)dt'\right\}.\label{eq:0-p}
\end{eqnarray}
Plugging the master equation Eq.(\ref{eq:maeq}) into above inequality,
we can obtain the lower bound of the purity, see Appendix
\ref{appendix c}.

In general, although it is difficult to calculate  exactly
 the integrals in above equation, it is still possible
to obtain a bound for the integrals, which would lead to a lower bound.
Noting that $|{\rho}_{ij}|\leq1$, $|{\rho}_{im}|\leq1$ and $|\exp(i\int_{0}^{t'}\omega_{nj}d\tau)|=1$,
we have
\begin{eqnarray}
1-p(t) & \leq & 4M\sum_{jm}(\left|\frac{\langle\Phi_{m}^{\bot}|\partial_{t}
|\Phi_{j}\rangle}{\omega_{mj}+i\Gamma_{m}}\right|\nonumber\\
&&+\int_{0}^{t}(A_{j}+B_{m}+C)\left|\frac{\langle\Phi_{m}^{\bot}
|\partial_{t'}|\Phi_{j}\rangle}{\omega_{mj}+i\Gamma_{m}}\right|dt'\nonumber\\
&&+\int_{0}^{t}\left|\frac{\partial}{\partial t'}\left(\frac{\langle\Phi_{m}^{\bot}|
\partial_{t''}|\Phi_{j}\rangle}{\omega_{mj}+i\Gamma_{m}}\right)\right|dt').\label{eq:1-p}
\end{eqnarray}
where
\begin{eqnarray*}
A_{j}&=&\sum_{k}|\langle\partial_{t}\Phi_{j}(t)|\Phi_{k}(t)\rangle|
+\sum_{n}|\langle\partial_{t}\Phi_{j}(t)|\Phi_{n}^{\bot}(t)\rangle|,\\
B_{m}&=&\sum_{k}|\langle\partial_{t}\Phi_{k}(t)|\Phi_{m}^{\bot}(t)
\rangle|+\sum_{n}|\langle\partial_{t}\Phi_{n}^{\bot}(t)|\Phi_{m}^{\bot}(t)\rangle|\\
&&+\sum_{n\neq m}|\langle\Phi_{n}^{\bot}(t)|\hat{\Gamma}|\Phi_{m}^{\bot}(t)\rangle|,\\
C&=&\frac{1}{M}\sum_{in}\left(|\langle\partial_{t}\Phi_{n}^{\bot}(t)|\Phi_{i}(t)
\rangle|+|\langle\partial_{t}\Phi_{i}(t)|\Phi_{n}^{\bot}(t)\rangle|\right)
\end{eqnarray*}
Since the sums on the right hand side of Eq.(\ref{eq:1-p})
are finite terms, the purity is very close  to 1 if each of the
terms can be omitted. Therefore, we can present stronger
conditions for ADFSs:
\begin{eqnarray*}
\left|\frac{\langle\Phi_{m}^{\bot}|\partial_{t}|\Phi_{j}\rangle}{\omega_{mj}
+i\Gamma_m}\right|\ll1, \ t\in[0,T],\\
\int_{0}^{T}(A_{jm}+B_{jm}+C)\left|\frac{\langle\Phi_{m}^{\bot}|\partial_{t'}|\Phi
_{j}\rangle}{\omega_{mj}+i\Gamma_m}\right|dt'\ll1,\\
\int_{0}^{T}\left|\frac{\partial}{\partial t'}
\left(\frac{\langle\Phi_{m}^{\bot}|\partial_{t'}|\Phi_{j}\rangle}
{\omega_{mj}+i\Gamma_m}\right)\right|dt'\ll1.
\end{eqnarray*}
where $T$ is the total evolution time for which the adiabatic approximation
is valid.

To relate the lower bound to the evolution time $T$, we may simplify
Eq.(\ref{eq:1-p}) by the following  estimations. By introduce the maximal
modulus of integrand in Eq.(\ref{eq:1-p}) for $t\in[0,T]$, we have
\begin{eqnarray}
1&-&p(T) \nonumber\\& \leq & 4M\sum_{jm}\left(\left|\frac{\langle\Phi_{m}^{\bot}|\partial_{t}
|\Phi_{j}\rangle}{\omega_{mj}+i\Gamma_m}\right|\right.\nonumber \\
&  & +\text{sup}_{t\in[0,T]}\left\{ (A_{jm}+B_{jm}+C)
\left|\frac{\langle\Phi_{m}^{\bot}|\partial_{t}|\Phi_{j}\rangle}
{\omega_{mj}+i\Gamma_m}\right|\right\} T\nonumber\\
&  & \left.+\text{sup}_{t\in[0,T]}\left\{ \left|\frac{\partial}{\partial t}
\left(\frac{\langle\Phi_{m}^{\bot}|\partial_{t}|\Phi_{j}\rangle}
{\omega_{mj}+i\Gamma_m}\right)\right|\right\} T\right).\label{eq:2-p}
\end{eqnarray}
Now we consider a quantum system defined by the parameterized Hamiltonian
$\hat{H}(s)$ and Lindblad operators $\hat{F}_{\alpha}(s)$, where
$s=t/T$, $t\in[0,T]$. Substituting $t=sT$ into Eq.(\ref{eq:2-p}),
we obtain the lower bound of the purity as a function of the total
evolution time $T$,
\begin{eqnarray*}
1&-&p(T) \nonumber\\& \leq & 4\frac{M}{T}\sum_{jm}\left(\left|\frac{\langle\Phi_{m}^{\bot}
|\partial_{s}|\Phi_{j}\rangle}{\omega_{mj}+i\Gamma_m}\right|\right.\\
&  & +\text{sup}_{s\in[0,1]}\left\{ (A_{j}+B_{m}+C)
\left|\frac{\langle\Phi_{m}^{\bot}|\partial_{s}
|\Phi_{j}\rangle}{\omega_{mj}+i\Gamma_m}\right|\right\} \\
&  & \left.+\text{sup}_{s\in[0,1]}\left\{ \left|\frac{\partial}{\partial t'}
\left(\frac{\langle\Phi_{m}^{\bot}|\partial_{s}
|\Phi_{j}\rangle}{\omega_{mj}+i\Gamma_m}\right)\right|\right\} \right).
\end{eqnarray*}
Since all  terms on the right-hand side of above equation can be arbitrarily
small as $T$ increases, the purity can reach 1 if $T$ is large
enough. We then arrive at the conclusion that the ADFS  is
valid for quantum systems which fulfills both the adiabatic
condition Eq.(\ref{eq:c2}) and the t-DFS conditions Eqs.(\ref{eq:Falpha}) and
(\ref{eq:Heff}) as long as $T$ is large enough.

\section{Shortcuts to adiabatic evolution of decoherence-free subspace}\label{shortcuts}

Equipped with the  theorem of ADFSs,  we are now ready to present STA dynamics for the ADFSs
with the language of the transitionless quantum driving method\cite{berry}. Consider an open quantum system
described by the master equation Eq.(\ref{eq:maeq}) and assume that there are t-DFSs $\mathcal
H_{\text{DFS}}(t)$ which is spanned by the common eigenstates of the Lindblad operators
$\{|\Phi_i (t)\rangle \}|_{i=1}^M$. Our purpose
is to drive the quantum state from the initial t-DFS into the target t-DFS without
any purity loss.
In the adiabatic approximation, the states driven by $\hat H_0(t)$ and $\{\hat F_\alpha(t)\}$ would
be
\begin{eqnarray*}
|\psi(t)\rangle=\sum_{i=1}^Mc_i(t)|\Phi_i(t)\rangle,
\end{eqnarray*}
with normalized parameters $c_i(t)$. In the transitionless quantum driving method adopted here, in order to realize the ADFS, we have to find
a Hamiltonian $\hat H(t)$ for the open quantum systems satisfying Eq.(\ref{shor}).

For the Hamiltonian and the Lindblad operators, the states must follow $\mathcal
H_{\text{DFS}}(t)$ exactly without any purity loss: there are no transition
between the t-DFS and the complementary
subspace. To obtain $\hat H(t)$, we notice that the
unitary operator $\hat U_{\text{DFS}}(t)$ is the solution of
\begin{eqnarray*}
i \partial_t \hat U_{\text{DFS}}(t)=\hat H_{\text{eff}}(t)\hat U_{\text{DFS}}(t),
\end{eqnarray*}
where
\begin{eqnarray}
\hat H_{\text{eff}}(t)=i \partial_t \hat U_{\text{DFS}}(t)\hat U_{\text{DFS}}^\dag(t).\label{H3}
\end{eqnarray}
Following the fact that $\hat U_{\text{DFS}}(t)$ can be parameterized by the common degenerate
eigenstates of Lindblad operators,
\begin{eqnarray}
\hat U_{\text{DFS}}(t)=\sum_{ij}u_{ij}(t)|\Phi_i (t)\rangle\langle\Phi_j(0)|.\label{U3}
\end{eqnarray}
where $\sum_{j}u_{ij}(t)u^*_{jk}(t)=\delta_{ik}$. Substituting Eq.(\ref{U3}) into Eq.(\ref{H3}), we immediately obtain,
\begin{eqnarray}
\hat H_{\text{eff}}(t)&=&i\sum_{i,j,l}\left(\partial_t u_{ij}(t)u^*_{jl}(t)|\Phi_i (t)\rangle\langle\Phi_l(t)|\right.\nonumber\\
&&\left.+u_{ij}(t)u^*_{jl}(t)|\partial_t\Phi_i (t)\rangle\langle\Phi_l(t)|\right)\nonumber\\
&&\equiv \hat H^0_{\text{eff}}(t)+\hat H_1(t).
\end{eqnarray}
Here $ \hat H^0_{\text{eff}}(t)$ is the operator mentioned in Eq.(\ref{heff}), satisfying Eq.(\ref{eq:Heff}).
At this time, the effective Hamiltonian can be written as
\begin{equation}
\hat{H}_{\text{eff}}(t)=\hat{H}(t)+\frac{i}{2}\sum_{\alpha}
\left(c_{\alpha}^{*}(t)\hat{F}_{\alpha}(t)-c_{\alpha}(t)\hat{F}_{\alpha}^{\dagger}(t)\right).
\end{equation}
with $\hat{H}(t)=\hat{H}_0(t)+\hat H_1(t)$. The elements of the off-diagonal block of $\hat H_1(t)$ can
be determined by considering  the t-DFSs condition Eq.(\ref{eq:Heff}), which reads
\begin{eqnarray}
\langle\Phi_n^\bot(t)|\hat H_1(t)|\Phi_k(t)\rangle
=i\langle\Phi_n^\bot(t)|\partial_t|\Phi_k(t)\rangle.\label{h1}
\end{eqnarray}
Notice that any  operator $\hat K(t)$ can be decomposed as two parts, $\hat K(t)=\hat K_N(t)+\hat K_Y(t)$ with $\hat K_N(t)$ denoting the off-diagonal parts and
$\hat K_Y(t)$ diagonal parts.  The off-diagonal parts
means it can be written as,
\begin{eqnarray}
\hat K_N(t)=\hat{P}(t)\hat{K}(t)\hat{Q}(t)+\hat{Q}(t)\hat{K}(t)\hat{P}(t),
\end{eqnarray}
with the projectors $\hat{P}(t)=\sum_{j}|\Phi_{i}(t)\rangle\langle\Phi_{i}(t)|$
on $\mathcal{H}_{\text{DFS}}(t)$ and $\hat{Q}(t)=\sum_{n}|\Phi_{n}^{\bot}(t)\rangle\langle\Phi_{n}^{\bot}(t)|$
on $\mathcal{H}_{\text{CS}}(t)$.
With this notation,  the off-diagonal block of the Hamiltonian of open quantum systems can be written as
\begin{eqnarray}
\langle\Phi_{j}(t)|&\hat{H}(t)&|\Phi_{n}^{\bot}(t)\rangle=
i\langle\Phi_n^\bot(t)|\partial_t|\Phi_k(t)\rangle\nonumber\\
&&-\frac{i}{2}\sum_{\alpha}c_{\alpha}(t)\langle\Phi_{j}(t)|
\hat{F}_{\alpha}(t)|\Phi_{n}^{\bot}(t)\rangle,\label{eq:HFx}
\end{eqnarray}
which is the very condition for t-DFSs discussed  in Ref.\cite{tdfs1}. In other words,
when the open quantum system is engineered by an coherent control   according to Eq.(\ref{eq:HFx}), the dynamics
of the quantum state lying in  $\mathcal H_{\text{DFS}}(t)$ is unitary. Hence, the quantum
state will follow the t-DFS from the initial DFS into a target DFS without
any fidelity loss.

The other parts of the Hamiltonian, i.e., the upper block and the
lower block of the Hamiltonian have no contribution to  shortcuts to ADFS. This can be shown in the definition of ADFS. I.e., by the ADFS, the quantum state can be    transported unitarily  from an  initial DFS into an target DFS,  but it is not necessary be transported  from an initial state to a
target state. The upper block of Hamiltonian, which can be rewritten as
\begin{eqnarray}
\hat H_D(t)=\hat{P}(t)\hat{H}(t)\hat{P}(t),
\end{eqnarray}
 is to induce the transition between the
bases of the t-DFS. Thus,  we do not need to take any special requirement on
the other parts of Hamiltonian. On the other hand, if we need to engineer the quantum
state into a special target state, the upper block of Hamiltonian have to be engineered. Since
the dynamics of quantum states in t-DFS is unitary, we can follow the standard procedure of the transitionless
quantum driving method to drive  the quantum state into an target state\cite{berry,sdfs}.

\section{Example}\label{sec3}

\subsection{The ADFSs for a Two-level System}

As an example, we consider a two-level system with ground state $|0\rangle$ and excited
state $|1\rangle$ coupled to both a broadband squeezed vacuum field and a coherent
control field $\Omega(t)$. Under the Markov approximation, the dynamics
of the two-level system is described by the following   master equation \cite{dfdi},
\begin{eqnarray}
\partial_t \rho(t)=-i[H_0(t),\rho(t)]+\mathcal L_D\rho(t).\label{liou}
\end{eqnarray}%
The Hamiltonian of two-level atom can be written as
\begin{eqnarray}
H_0(t)=\Omega(t)\ket{0}\bra{1}+h.c..\label{Ham}
\end{eqnarray}%
The dissipator caused by the coupling to the squeezed vacuum is
\begin{eqnarray}
\mathcal L_D \rho(t)&=&\gamma \cosh^2(r)\left(\sigma_+\rho(t)\sigma_-
-\frac{1}{2}\{\sigma_+\sigma_-\rho(t)\}\right)\nonumber\\&&+\gamma
\sinh^2(r)\left(\sigma_-\rho(t)\sigma_+
-\frac{1}{2}\{\sigma_-\sigma_+\rho(t)\}\right)\nonumber\\
&&+\gamma\sinh(r)\cosh(r) \exp(-i \theta) \sigma_-\rho(t)\sigma_-\nonumber\\
&&+\gamma \sinh(r)\cosh(r) \exp(i \theta) \sigma_+\rho(t)\sigma_+,\label{maeq2}
\end{eqnarray}%
where $r$ is the squeezing strength and $\theta$ is the squeezing phase,
$\sigma_-$ ($\sigma_+$) is the lowing (raising) operator, $\gamma$ is
the spontaneous decay rate. In Eq.(\ref{maeq2}), we have assumed that
the vacuum squeezing field is perfect. If we redefine the decoherence
operator as follows,
\begin{eqnarray}
L=\cosh(r)\exp(-i\theta/2)\sigma_-+\sinh(r)\exp(i\theta/2)\sigma_+,\label{Lindblad}
\end{eqnarray}%
the dissipator could be transformed into the Lindblad form,
\begin{eqnarray}
\mathcal L_D \rho(t)=\frac{\gamma}{2}\left(2 L \rho(t) L^\dag-\{L^\dag L, \rho(t) \}\right).
\end{eqnarray}%

\begin{figure}
\includegraphics[scale=0.45]{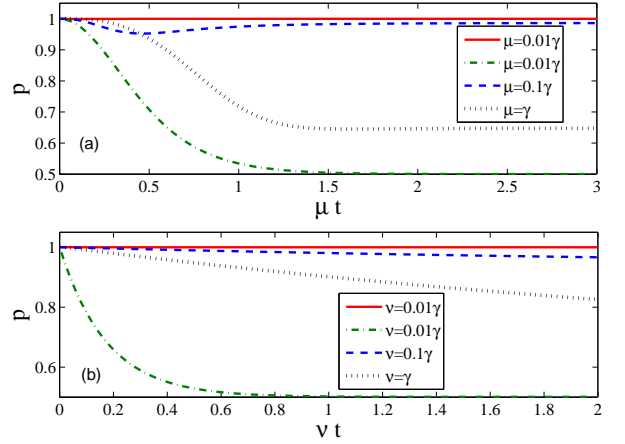}
\centering
\caption{(Color online) (a) The purity  $p(t)$  versus dimensionless parameter  $\mu t$  (in units of $\pi$) with parameters $\nu=0$, $\mu=0.01\gamma$ (the red solid line), $\mu=0.1\gamma$ (the blue dash line) and $\mu=\gamma$ (the black dot line);  (b) The purity   versus dimensionless parameter  $\nu t$ (in units of $\pi$) with parameters $\mu=0$, $\nu=0.01\gamma$ (the red solid line), $\nu=0.1\gamma$ (the blue dash line) and $\nu=\gamma$ (the black dot line). The results are obtained by calculating the master equation without the coherent control field $\Omega(t)$ (the green dot dash line) and with the coherent control field (the other lines). }\label{Pt}
\end{figure}

\begin{figure}
\centering
\includegraphics[scale=0.45]{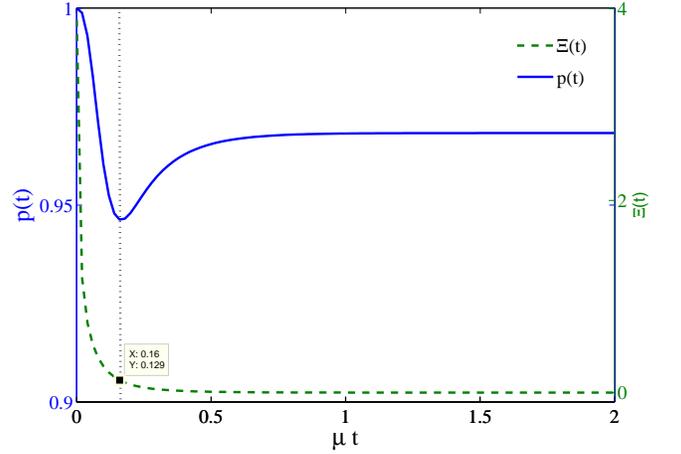}
\caption{(Color online) The purity   $p(t)$ and the adiabatic condition
$\Xi(t)$ versus dimensionless parameters $\mu t$  (in units of $\pi$) with $\nu=0$ and $\mu=0.1\gamma$.
The initial state is same as  in in FIG.\ref{Pt} (a).}\label{2}
\end{figure}

According to the  theorem for ADFSs mentioned in Sec.\ref{sub:adfs},
two independent conditions are required by the ADFSs. \emph{the t-DFSs
condition} and \emph{the adiabatic condition}. On   one hand, a subspace spanned by
$\mathcal H_{\text{DFS}}= \{\ket{\phi}\}$ is a t-DFS of
the two-level system, if $\ket{\phi}$
is the eigenvector of the Lindblad operator $L$. The Lindblad
operator $L$ (Eq.(\ref{Lindblad})) gives two nonorthogonal eigenstates,
\begin{eqnarray}
\ket{\phi_1}=\frac{1}{n}\left(\sqrt{\sinh(r)}\exp(i\theta/2)\ket{0}+\sqrt{\cosh{(r)}}\ket{1}\right),\label{phi1}\\
\ket{\phi_2}=\frac{1}{n}\left(-\sqrt{\sinh(r)}\exp(i\theta/2)\ket{0}+\sqrt{\cosh{(r)}}\ket{1}\right),
\end{eqnarray}%
with eigenvalues $\lambda_1=\sqrt{\sinh(r)\cosh(r)}$ and $\lambda_2=-\sqrt{\sinh(r)\cosh(r)}$,
in which $n=\sinh(r)+\cosh(r)$ is normalizing factor. Anyone of the eigenstates can
be the basis of subspace $\mathcal H_{\text{DFS}}$. Without losing the generality, we choose
$\ket{\phi_1}$  as the basis of $\mathcal H_{\text{DFS}}$. Then
the basis of the orthogonal complement space can be  determined, which reads
\begin{eqnarray}
\ket{\phi^\bot}=\frac{1}{n}\left(\sqrt{\cosh(r)}\exp(i\theta/2)\ket{0}-\sqrt{\sinh{(r)}}\ket{1}\right).\label{phio}
\end{eqnarray}%
At the same time, the two-level system have to be
engineered according to Eq.(\ref{eq:Heff}).  Taking the Hamiltonian Eq.(\ref{Ham})
and the Lindblad operator Eq.(\ref{Lindblad}) into
the effective Hamiltonian $H_{\text{eff}}^0(t)$ and considering the condition Eq.(\ref{eq:Heff}), we are
able to write the accurate function of the coherent control field as
\begin{eqnarray}
\Omega(r,\theta)&=&\frac{i\gamma\exp(-r-i \theta)\sqrt{\sinh(r)\cosh(r)}}{2}.\label{cont}
\end{eqnarray}%
We engineer the surroundings of the two-level atom from the vacuum field to the
squeezed vacuum field by means of engineering reservoir technology
(the incoherent control program)\cite{re1,re2},
which results in the time dependence of the squeezed parameters. For simplicity,
both the squeezed strength and the squeezed  phase are set to depend on time linearly,
i.e., $r=r_0+\mu t$ and $\theta=\theta_0+\nu t$ with initial squeezed parameters $r_0$
and $\theta_0$. By combining the incoherent control
Eq.(\ref{Lindblad}) with the coherent control Eq.(\ref{cont}), the instantaneous
DFS $\mathcal H_{\text{DFS}}$ is dynamically stable.

\begin{figure}
\centering
\includegraphics[scale=0.5]{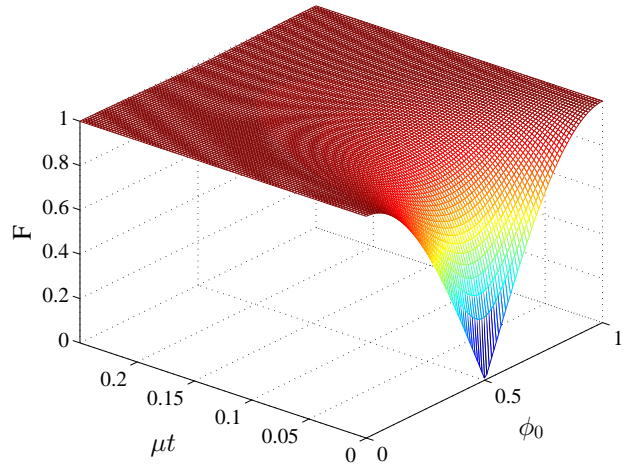}
\caption{(Color online) The fidelity between the quantum states $\rho(t)$ and
$\rho_{\text{DFS}}(t)=|\phi_1(t)\rangle \langle\phi_1(t)|$ versus $\mu t$ (in units of $\pi$)
and $\varphi$ (in units of $\pi$)  with $\mu=\nu=0.1\gamma.$}\label{3}
\end{figure}

On the other hand, when the Lindblad operator Eq.(\ref{Lindblad}) is engineered by
incoherent control continuously, the t-DFS $\mathcal H_{\text{DFS}}$ is time-dependent.
To compel the quantum state of the two-level system to follow the track of the t-DFS,
the adiabatic condition Eq.(\ref{eq:maxnj}) have to be satisfied. Taking Eqs.(\ref{Lindblad}),
(\ref{phi1}) and (\ref{phio}) into Eq.(\ref{eq:maxnj}), the adiabatic condition reads
\begin{eqnarray}
\Xi=\left|\frac{4i\left(\mu+i\nu \,\sinh\left(r\right)\cosh\left(r\right)\right)}
{\gamma\sqrt{\sinh\left(r\right)\cosh\left(r\right)}\left(\sinh\left(3r\right)
+\cosh\left(3r\right)\right)}\right|,\label{xi1}
\end{eqnarray}%
which characterizes  the
broken adiabaticity. In fact, it can be verified that
the spectrum of the effective Hamiltonian is degenerate, i.e., $\hat H_{\text{eff}}(t)=\bf{0}$. But
the adiabatic condition is still available as shown in  Eq.(\ref{xi1}), which results from the
contribution of the decoherence process.

\begin{figure*}
\centering
\begin{minipage}[!htbp]{\linewidth}
\includegraphics[scale=0.8]{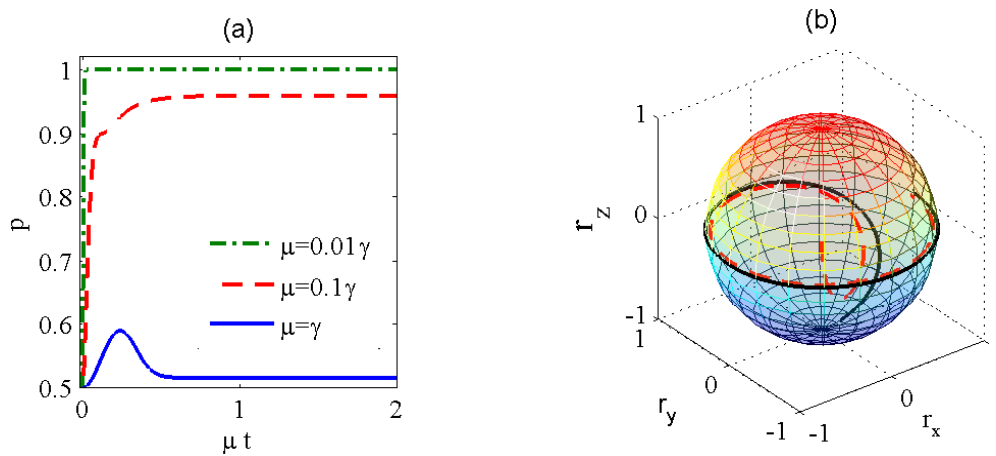}
\caption{(Color online) (a) The purity  versus dimensionless parameters
$\mu t$ (in units of $\pi$) with $\mu=\nu=0.01\gamma$ (the green dot dash line),
$\mu=\nu=0.1\gamma$ (the red dash line), and $\mu=\nu=\gamma$ (the blue solid line).
(b) The Bloch vectors for $\rho(t)$ (the red dash line) and $\rho_{\text{DFS}}(t)=|\phi_1(t)\rangle \langle\phi_1(t)|$
(the black solid line) with $\mu=\nu=0.1\gamma$.  In the figures, the initial states are chosen as the maximally mixed state.}\label{4}
\end{minipage}%
\end{figure*}

The numerical results of purity versus dimensionless parameters $\mu t$ and $\nu t$ were
illustrated in FIG. \ref{Pt}. The initial states are prepared in the initial
t-DFS $\mathcal H_{\text{DFS}}(0)$, i.e., $\rho(0)=|\phi_1(0)\rangle\phi_1(0)|$ with
$r_0=0$ and $\theta_0=0$  in FIG. \ref{Pt} (a) and with $r_0=2\pi$ and $\theta_0=0$  in FIG. \ref{Pt} (b).
According to Eq.(\ref{xi1}), the adiabatic condition can be
held, if the parameters $\mu$ and $\nu$ is far less than the spontaneous decay rate $\gamma$.
In other words, the smaller $\mu$ and $\nu$
are, the larger the purity is at the end of evolution,
 which is verified by FIG. \ref{Pt}. Besides, we also consider the effect
of the coherent control field $\Omega(t)$ on the purity. When the coherent control field is absent
(green dot dash line), the quantum state becomes a maximally mixed state in minimal
time scale, even the parameters $\mu$ and $\nu$ are very small. Therefore,
the cooperation between the coherent control and the incoherent control is significant
for the ADFSs.

It can be observed in FIG.\ref{Pt} (a) that the dependence of the purity on
the parameter $\mu t$ is not monotonic: The purity loses evidently
at the beginning of the engineering stage. After that, the purity
increases and tends to a stable value. This is due to the competition between the broken
adiabaticity   and the decay effect from the complementary subspace to the t-DFS.
To illustrate  the result in detail, the purity $p(t)$ and the adiabatic condition $\Xi(t)$ as a function of the dimensionless
parameter $\mu t$ are plotted in FIG.\ref{2}, in which the change rate of the squeezed parameters  are chosen as
$\mu=0.1\gamma$ and $\nu=0$. The initial squeezed parameters are set to be $r_0=0$ and $\theta_0=0$.
It is convenient to break the abscissa axis into two intervals. The boundary  is the
black dot line in FIG.\ref{2}, where the purity is minimal and $\Xi\doteq 0.129$.
When the adiabatic condition is broken,  the quantum state can not follow the t-DFS
(the interval at the left side of the black dot line), so that the purity decreases.
This can also be found in Eq.(\ref{xi1}). With the evolution, $\Xi$ decays into the region where the
adiabatic condition is satisfied (at the right side of the block dot line). At this time, the decay effect is
stronger than the broken adiabaticity. Since the decoherence process induces the transition from
the complementary subspace into the t-DFS, the fidelity of the quantum state increases gradually  until it reaches a
stable value, which can be explained as continuous Zeno measurements projecting onto the t-DFSs\cite{zeno}.
Form FIGs.\ref{Pt} and \ref{2}, we may conclude that the adiabatic condition presented here is available
to judge whether the adiabaticity is broken for the ADFSs.

From the observation from FIG.\ref{2}, one may wonder that if we do not prepare the initial state in
$\mathcal H_{\text{DFS}}(0)$, whether the quantum state can be drawn back into $\mathcal H_{\text{DFS}}(t)$ due
to the decay effect. The numerical results have been shown in Figures\ref{3} and \ref{4}, in which
both pure and mixed initial states are considered. On one side, in FIG.\ref{3}, the initial state are chosen as
a pure state $|\psi\rangle=\sin\phi_0|0\rangle+\cos\phi_0|1\rangle$. We consider the dependence of
the fidelity between the quantum states $\rho(t)$ and $\rho_{\text{DFS}}(t)=|\phi_1(t)\rangle\langle\phi_1(t)|$
on the parameters $\mu t$ and $\varphi_0$. The change rate of the squeezed parameters
are chosen to be $\mu=\nu=0.1\gamma$. The numerical result illustrates that, due to the decay, the fidelity increases with the evolution. The final fidelity approaches to 1,  even though the initial state does not be prepared in
$\mathcal H_{\text{DFS}}(0)$ exactly. On the other side, we also consider the case where the initial state
is the maximally  mixed state, i.e.,
\begin{eqnarray}
\rho(0)=\frac{|0\rangle\langle0|+|1\rangle\langle1|}{2}.
\end{eqnarray}%
As shown in FIG. \ref{4}(a), the purity  increase with the
parameter $\mu t$ which can also be explained as a result of the decay effect.
What is more, we can observe that the final purities are sensitive with the
adiabatic condition. When the adiabatic condition is satisfied, the quantum state is drawn back into t-DFSs
perfectly (the green dot dash line). On the contrary, when the adiabatic condition is broken, the
system can not be derived into the t-DFS (the blue solid line). In other words, the quantum state will be drawn back into t-DFSs if the
adiabatic condition is satisfied. For illustrating the conclusion more intuitively, we plot the Bloch vectors of
the quantum states $\rho$ and $\rho_{\text{DFS}}(t)$ in FIG. \ref{4}(b) in case of $\mu=0.1\gamma$.
It can be observed that the Bloch vector of $\rho(t)$ starts from the center of the Bloch sphere
(the maximally mixed state), which is illustrated by the red dash line in FIG. \ref{4}(b). With the evolution, the quantum state
$\rho(t)$ are approaching to $\rho_{\text{DFS}}(t)$, until they have a stable distance in the Bloch sphere. According to the above observation, we conclude that the requirement on the preparation of the initial
state can be relaxed. When the adiabatic condition is satisfied, we still can engineer the quantum state of quantum
systems into the target subspace for arbitrary  initial states.

\subsection{protocol for shortcuts to ADFSs}

Let us now apply the transitionless quantum driving method by
taking the interaction picture Hamiltonian Eq.(\ref{Ham}) as the reference Hamiltonian
$\hat H_0(t)$. Choosing $\{|\phi_1(t)\rangle, \,\phi^\bot(t)\rangle\}$ given by Eqs.(\ref{phi1})
and (\ref{phio}) as the bases of the t-DFS and the
complementary subspace, the driving Hamiltonian Eq. (\ref{h1}) becomes in this case,
\begin{eqnarray}
H_1(t)=\Omega'(t)\ket{0}\bra{1}+h.c.,\label{Ham1}
\end{eqnarray}%
with the coherent control field
\begin{eqnarray}
\Omega'(r,\theta)&=&\frac{i\exp(-r-i \theta)(\mu-i\nu \sinh(r)\cosh(r))}
{2\sqrt{\sinh(r)\cosh(r)}}.\nonumber\\\label{cont}
\end{eqnarray}%

\begin{figure*}
\centering
\begin{minipage}[!htbp]{\linewidth}
\includegraphics[scale=0.6]{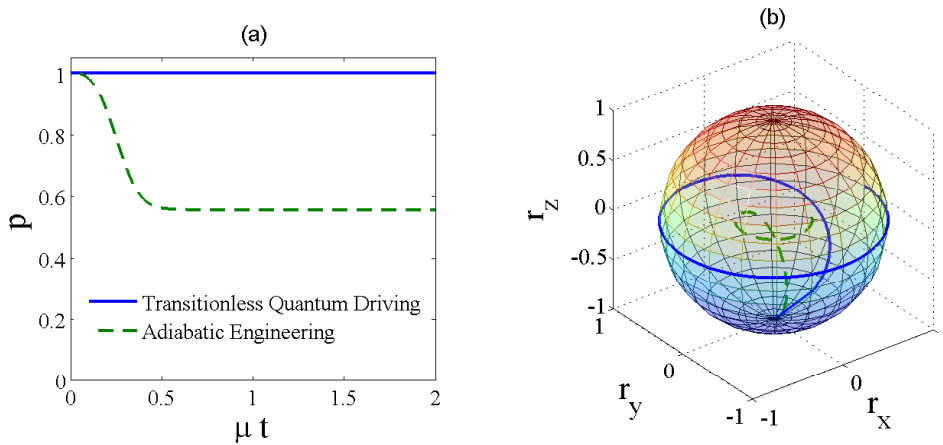}
\caption{(Color online) (a) The purity  versus dimensionless parameters
$\mu t$ (in units of $\pi$) with $\mu=\nu=\gamma$ and $o=0.01$.
(b) The Bloch vectors for the quantum state of the two-level system.
In the figures, the results are obtained for the cases where the two-level
system is driven by the Hamiltonian $\hat H(t)$ (blue solid lines) and
$\hat H_0(t)$ (green dash lines).}\label{5}
\end{minipage}%
\end{figure*}

We implement $\hat H(t)=\hat H_0(t)+\hat H_1(t)$ as the Hamiltonian of the two-level system to illustrate
practicability of the shortcuts to ADFSs.  The numerical results are illustrated in FIG.\ref{5}. Both the squeezed
parameter and the squeezed phase are set to depend on time linearly,
\begin{eqnarray}
r(t)=\mu t+o,\ \theta=\nu t,
\end{eqnarray}%
where $o$ is an extremely small constant. In FIG. \ref{5} (a), we consider the purity as a function
of  $\mu t$ with $\mu=\nu=\gamma$. The initial state of the two-level system is prepared in
$\mathcal H_{\text{DFS}}(0)$. As shown in the adiabatic condition Eq.(\ref{xi1}), the purity
cannot hold  in this case, when we use $\hat H_0(t)$ as the Hamiltonian
of the two-level system. The purity of the quantum state decreases rapidly, since the adiabatic
condition is not fulfilled (the green dash line in FIG. \ref{5} (a)). On the other hand, if the transitionless
quantum driving method is used, the system remains in the t-DFS without any
purity loss (the blue solid line in FIG. \ref{5} (a)). We also plot
the Bloch vectors for both the ADFSs and the transitionless quantum
driving in FIG. \ref{5} (b), which   confirms  our conclusion obtained in FIG. \ref{5}
(a). As a result, the protocol  proposed in Sec. \ref{shortcuts} is the shortcuts to
adiabaticity for the ADFSs, which can be used to accelerate
the ADFSs  definitely.

\section{Conclusion}\label{conclusion}
Starting from the definition of dynamical stable DFSs, we have presented an adiabatic theorem for an open system in a time-dependent  decoherence-free subspaces. An adiabatic condition to guarantee the quantum system in the time-dependent decoherence-free subspace is also presented.

The adiabatic theorem contains  two independent conditions, i.e., the condition for t-DFSs
and the condition for adiabatic evolution.
The first condition states that both  the coherent evolution(governed by the Hamiltonian)  and  the incoherent evolution(governed by the Lindbladian) are important to enable the DFSs dynamically stable.
The second condition is  the adiabatic condition
Eq.(\ref{eq:adicond}) (also see Eqs.(\ref{cond2})), which  suggests  that if the variation of the effective frequency of the Lindblad operators is smaller than  the transition frequencies $|\omega_{ni}+i\Gamma_n|$,
the quantum state of open quantum system would keep in the t-DFS.

To quantify the probability  of the system in the ADFSs, we derive a  lower bound for  the probability. Further, we have also proposed a STA
protocol(i.e., the shortcuts-to-adiabaticity for open systems) for ADFSs.
Following the transitionless quantum driving method
for closed quantum systems, we show that the quantum state can keep in  the t-DFS without any purity loss by adjusting the coherent evolution.

\emph{This work is supported by NSF of China under Grant Nos. 11605024, 11534002, 61475033}.

\appendix

\begin{widetext}

\section{ The Adiabatic Condition Eq.(\ref{eq:maxnj})}\label{sec:Appendix-A}

In this part of appendix, we show that Eq.(\ref{eq:maxnj}) is the
condition approved for ADFSs. Firstly, we introduce a transformation
operator
\begin{eqnarray}
\hat T(t) & = & \sum_j\exp\left(i\int_{0}^{t}\langle\hat{H}_{\text{eff}}^{0}(\tau)\rangle_{i}d\tau\right)
|\Phi_{i}(0)\rangle\langle\Phi_{i}(t)|\nonumber\\
 &  & +\sum_{n}\exp\left(i\int_{0}^{t}(\langle\hat{H}_{\text{eff}}^{0}(\tau)
 \rangle_{n}+i\langle\hat{\Gamma}(\tau)
 \rangle_{n})d\tau\right)|\Phi_{n}^{\bot}(0)\rangle\langle\Phi_{n}^{\bot}(t)|,\label{eq:U}
\end{eqnarray}
where $\langle\hat{H}_{\text{eff}}^{0}(\tau)\rangle_{i}=\langle\Phi_{i}(\tau)
|\hat{H}_{\text{eff}}^{0}(\tau)|\Phi_{i}(\tau)\rangle$,
$\langle\hat{H}_{\text{eff}}^{0}(\tau)\rangle_{n}=\langle\Phi_{n}^\bot(\tau)
|\hat{H}_{\text{eff}}^{0}(\tau)|\Phi_{n}^\bot(\tau)\rangle$, and
$\langle\hat{\Gamma}(\tau)\rangle_{n}=\sum_\alpha\langle\Phi_{n}^\bot(\tau)|
(\hat{F}_{\alpha}^\dag(\tau)-c_\alpha^*(\tau))(\hat{F}_{\alpha}(\tau)
-c_\alpha (\tau))|\Phi_{n}^\bot(\tau)\rangle/2$.
Besides, the inverse transformation operator can be written as
\begin{eqnarray}
\hat T^{-1}(t) & = & \sum_j\exp\left(-i\int_{0}^{t}\langle\hat{H}_{\text{eff}}^{0}(\tau)\rangle_{i}d\tau\right)
|\Phi_{i}(t)\rangle\langle\Phi_{i}(0)|\nonumber\\
 &  & +\sum_{n}\exp\left(-i\int_{0}^{t}(\langle\hat{H}_{\text{eff}}^{0}(\tau)
 \rangle_{n}+i\langle\hat{\Gamma}(\tau)
 \rangle_{n})d\tau\right)|\Phi_{n}^{\bot}(t)\rangle\langle\Phi_{n}^{\bot}(0)|
\end{eqnarray}
We transform the density matrix $\hat \rho(t)$ into a rotating frame by this transformation
 operators, i.e. $\bar{\rho}(t)=\hat T(t)\hat{\rho}(t)\hat T^{-1}(t)$. At this time,
the master equation Eq.(\ref{eq:maeq}) reads
\begin{eqnarray*}
\partial_{t}\bar{\rho}(t) & = & \bar{\mathcal{L}}(\bar{\rho}),
\end{eqnarray*}
where
\begin{equation}
\bar{\mathcal{L}}(\bar{\rho})=-i[\bar{H}+\bar{G},\bar{\rho}]+\sum_{\alpha}\left(\bar{F}_{\alpha}\bar{\rho}
\bar{F}_{\alpha}^{\dagger}-\frac{1}{2}\{\bar{F}_{\alpha}^{\dagger}\bar{F}_{\alpha},\bar{\rho}\}\right)\label{eq:liou2}
\end{equation}
with $\bar{H}=\hat T\hat{H}_0\hat T^{-1}$, $\bar{F}_{\alpha}=\hat T\hat{F}_{\alpha}\hat T^{-1}$
and $\bar{G}=i\partial_{t}\hat T\cdot\hat  T^{-1}$. By defining the new
Lindblad operator as $\tilde{F}_{\alpha}=\bar{F}_{\alpha}-c_{\alpha}$,
the decoherence terms in Eq.(\ref{eq:liou2}) can be rewritten as
\begin{equation}
\bar{\mathcal{L}}(\bar{\rho})=-i[\bar{H}_{\text{eff}}+\bar{G},\bar{\rho}]
+\sum_{\alpha}\left(\tilde{F}_{\alpha}\bar{\rho}\tilde{F}_{\alpha}^{\dagger}
-\frac{1}{2}\{\tilde{F}_{\alpha}^{\dagger}\tilde{F}_{\alpha},\bar{\rho}\}\right),\label{eq:liou3}
\end{equation}
in which $\bar{H}_{\text{eff}}=\hat T\hat{H}_{\text{eff}}^{0}\hat T^{\dagger}$
with the effective Hamiltonian mentioned in Eq.(\ref{eq:Heff}).
Hence, according to  the instantaneous  DFSs condition (see Eqs.(\ref{eq:Falpha}) and (\ref{eq:Heff})),
the operators used in Eq.(\ref{eq:liou3}) imply the following
properties: (1) $\tilde{F}_{\alpha}|\Phi_{i}(0)\rangle=0,$ $\forall i$;
(2) $\langle\Phi_{n}^{\bot}(0)|\bar{H}_{\text{eff}}|\Phi_{j}(0)\rangle=0,\,\forall n,\,j$.

With the definition of the dynamical stable DFSs, we consider
the time derivative  of the purtiy $p(t)$ of the quantum state $\hat{\rho}$,
\[
\partial_{t}p(t)=2\text{Tr}\{\hat{\rho}\partial_{t}\hat{\rho}\}=2\text{Tr}\{\bar{\rho}\partial_{t}\bar{\rho}\}.
\]
The quantum state $\bar{\rho}$ can be decomposed into three parts,
\begin{equation}
\bar{\rho}=\bar{\rho}_{D}+\bar{\rho}_{N}+\bar{\rho}_{C},\label{eq:derho}
\end{equation}
 where $\bar{\rho}_{D}$ ($\bar{\rho}_{C}$) is diagonal block corresponding to
DFSs (complementary subspaces) and $\bar{\rho}_{N}$ is off-diagonal
part. It is straight forward to obtain following commutation relations,
\[
[\bar{\rho}_{D},\bar{\rho}_{N}]\neq0,\ [\bar{\rho}_{C},\bar{\rho}_{N}]\neq0,\ [\bar{\rho}_{D},\bar{\rho}_{C}]=0.
\]
By introducing the projectors $\bar{P}=\sum_{j}|\Phi_{i}(0)\rangle\langle\Phi_{i}(0)|$
on $\mathcal{H}_{\text{DFS}}(0)$ and $\bar{Q}=\sum_{n}|\Phi_{n}^{\bot}(0)\rangle\langle\Phi_{n}^{\bot}(0)|$
on $\mathcal{H}_{\text{CS}}(0)$, we can rewrite the operators as $\bar{\rho}_{D}=\bar{P}\bar{\rho}\bar{P}$,
$\bar{\rho}_{C}=\bar{Q}\bar{\rho}\bar{Q}$, and $\bar{\rho}_{N}=\bar{P}\bar{\rho}\bar{Q}+\bar{Q}\bar{\rho}\bar{P}$.
When the total operation time $T$ goes
to infinity, the adiabatic evolution is achieved reliably \cite{adi2}.
In other words, in the adiabatic approximation, $\lim_{T\rightarrow\infty}\bar{\rho}_{D}=\bar{\rho}$
and $\lim_{T\rightarrow\infty}\bar{\rho}_{C}=\mathbf{0}$, $\lim_{T\rightarrow\infty}\bar{\rho}_{N}=\mathbf{0}$.
Therefore, the purity can be expressed approximately as
\[
\partial_{t}p(t)=2\lim_{T\rightarrow\infty}
\text{Tr}\{\bar{\rho}_{D}\partial_{t}\bar{\rho}\}.
\]
Taking Eq.(\ref{eq:liou3}) and Eq.(\ref{eq:derho}) into above equation,
we immediately obtain
\begin{eqnarray}
\partial_{t}p(t) & = & 2\lim_{T\rightarrow\infty}\text{Tr}\left\{\bar{\rho}_{D}(-i[\bar{H}_{\text{eff}}+\bar{G},
\bar{\rho}_{D}+\bar{\rho}_{N}+\bar{\rho}_{C}]+\sum_{\alpha}\left(\tilde{F}_{\alpha}
(\bar{\rho}_{D}+\bar{\rho}_{N}+\bar{\rho}_{C})\tilde{F}_{\alpha}^{\dagger}
-\frac{1}{2}\{\tilde{F}_{\alpha}^{\dagger}\tilde{F}_{\alpha},\bar{\rho}_{D}
+\bar{\rho}_{N}+\bar{\rho}_{C}\}\right))\right\}\nonumber \\
 & = & 2\lim_{T\rightarrow\infty}\text{Tr}\{\bar{\rho}_{D}(-i[\bar{G},\bar{\rho}_{N}]
 +\sum_{\alpha}\tilde{F}_{\alpha}\bar{\rho}_{C}\tilde{F}_{\alpha}^{\dagger})\},\label{eq:tpur}
\end{eqnarray}
where the following relations have been used: $\bar{P}\bar{H}_{\text{eff}}\bar{Q}=\mathbf{0}$,
$\tilde{F}_{\alpha}\bar{P}=\mathbf{0}$, $\bar{P}\tilde{F}_{\alpha} ^\dag=\mathbf{0}$,
and $[\bar{\rho}_{D},\bar{\rho}_{C}]=\mathbf{0}$. It's worth noting that the second term
of right-hand side of above equation describes the population transition from
the complementary subspace to the t-DFS. Further, the population
in the complementary subspace is very small under the adiabatic approximation.
Therefore, the contribution of this term on the evolution of the purity
is so negligible that this term can be ignored. Thus we have
\begin{eqnarray*}
\partial_{t}p(t) & = & 2\lim_{T\rightarrow\infty}\text{Tr}\{\bar{\rho}_{D}(-i[\bar{G},\bar{\rho}_{N}])\}\\
 & = & -2i\lim_{T\rightarrow\infty}\text{Tr}\{\bar{\rho}(\bar{\rho}_{D}\bar{G}_{R}-\bar{G}_{L}\bar{\rho}_{D})\}\\
 & = & -2i\lim_{T\rightarrow\infty}\langle\bar{\rho}_{D}\bar{G}_{R}-\bar{G}_{L}\bar{\rho}_{D}\rangle,
\end{eqnarray*}
where $\bar{G}_{R}=\bar P \bar G \bar Q$ and $\bar{G}_{L}=\bar Q \bar G \bar P$. The time derivative  of the purity can be rewritten in terms of the bases of $\mathcal H_{\text{DFS}}(t)$
and  $\mathcal H_{\text{CS}}(t)$ as
\begin{eqnarray}
 \partial_{t}p(t)
 & = & -2i\sum_{i,j,n}\bar{\rho}_{ij}^{D}\left(\bar{\rho}_{jn}^{N}\langle\Phi_{n}^{\bot}(t)|
 \partial_{t}|\Phi_{i}(t)\rangle\exp\left(i\int_{0}^{t}(\omega_{ni}(\tau)+i\langle\hat{\Gamma}(\tau)
 \rangle_{n})d\tau\right)\right.\\
 &  & \left.+\langle\partial_{t}\Phi_{j}(t)|\Phi_{n}^{\bot}(t)\rangle\bar{\rho}_{ni}^{N}
 \exp\left(-i\int_{0}^{t}(\omega_{ni}(\tau)+i\langle\hat{\Gamma}(\tau)
 \rangle_{n})d\tau\right)\right).\label{pt1}
\end{eqnarray}
To obtain above expression, we have used the fact that
\begin{eqnarray*}
\bar{\rho}_{jn}^{N}&=&\langle\Phi_{j}(0)|\hat T\hat{\rho}\hat T^{-1}|\Phi_{n}^{\bot}(0)\rangle\nonumber\\
&=&\langle\Phi_{j}(t)|\hat{\rho}(t)
|\Phi_{n}^{\bot}(t)\rangle\exp\left(-i\int_{0}^{t}(\omega_{nj}(\tau)+i\langle\Gamma(\tau)\rangle_{n})d\tau\right).
\end{eqnarray*}
Since the first term in Eq.(\ref{pt1}) is conjugate to the second term,
we can rewrite above expression as
\begin{eqnarray*}
\partial_{t}p(t)= 4\text{Re}\left\{\sum_{i,j,n}\bar{\rho}_{ij}^{D}\bar{\rho}_{jn}^{N}\langle\Phi_{n}^{\bot}
 (t)|\partial_{t}|\Phi_{i}(t)\rangle\exp\left(i\int_{0}^{t}(\omega_{ni}(\tau)+i\langle\hat{\Gamma}(\tau)
 \rangle_{n})d\tau\right)\right\}
\end{eqnarray*}
It is convenient to consider a dimensionless parameter $s=t/T$, where $T$
is total evolution time. We rewrite above expression as
\begin{eqnarray*}
 \partial_{s}p(s)
= 4\text{Re}\left\{\sum_{i,j,n}\bar{\rho}_{ij}^{D}\bar{\rho}_{jn}^{N}\langle\Phi_{n}^{\bot}(s)
 |\partial_{s}|\Phi_{i}(s)\rangle\exp\left(-iT\int_{0}^{s}(\omega_{ni}(s')+i\langle\hat{\Gamma}(s')
 \rangle_{n})ds'\right)\right\}.
\end{eqnarray*}
Integrating by parts, we have
\begin{eqnarray}
p(s)-1\doteq4\text{Re}\left\{ \sum_{ijm}\left(\frac{i}{T}\bar{\rho}_{ij}\bar{\rho}_{jn}\frac{\langle\Phi_{n}^{\bot}(s)
|\partial_{s}|\Phi_{i}(s)\rangle}{\omega_{ni}(s)+i\langle\hat{\Gamma}(s)
 \rangle_{n}}\exp\left(-iT\int_{0}^{s}
(\omega_{ni}(s')+i\langle\hat{\Gamma}(s')\rangle_{n})ds'\right)\right)\right\} ,
\end{eqnarray}
where we have omitted the terms higher than the second order of $T^{-1}$. Therefore,
in case of $$\left|\frac{4\langle\Phi_{n}^{\bot}(t)|\partial_{t}|\Phi_{i}(t)\rangle}
{\omega_{ni}(t)+i\langle\hat{\Gamma}(t)
 \rangle_{n})}\right|\ll1\ \forall n,i,$$
the purity very approaches to 1. In other
words, the adiabatic DFS can be acchieved, if the condition
\[
\text{Max}_{\{n,i\}}\left|\frac{4\langle\Phi_{n}^{\bot}(t)|\partial_{t}|\Phi_{i}(t)\rangle}
{\omega_{ni}(t)+i\langle\hat{\Gamma}(t)
 \rangle_{n}}\right|\ll1
\]
is satisfied, which is the adiabatic condition Eq.(\ref{eq:maxnj}) presented in Sec. \ref{sub:adfs}.

\section{The Derivation of Eq.(\ref{eq:adicond})}\label{appendix b}

We start with the eigen-equation of Lindblad operators Eq.(\ref{eq:Falpha}).
Taking the derivative of time on the eigen-equation, we have
\[
\partial_{t}\hat{F}_{\alpha}(t)|\Phi_{i}(t)\rangle+\hat{F}_{\alpha}(t)\partial_{t}|\Phi_{i}(t)\rangle
=\partial_{t}c_{\alpha}(t)|\Phi_{i}(t)\rangle+c_{\alpha}(t)|\Phi_{i}(t)\rangle.
\]
Then we act $\langle\Phi_{n}^{\bot}(t)|$ from the left to obtain
\begin{equation}
\langle\Phi_{n}^{\bot}(t)|\partial_{t}\hat{F}_{\alpha}(t)|\Phi_{i}(t)\rangle+\langle\Phi_{n}^{\bot}(t)|\hat{F}_{\alpha}(t)\partial_{t}
|\Phi_{i}(t)\rangle=c_{\alpha}(t)\langle\Phi_{n}^{\bot}(t)|\partial_{t}|\Phi_{i}(t)\rangle.\label{eq:b2}
\end{equation}
 We plug a complete set into the second term of above equation and
consider the eigen-equation of Lindblad operators, to obtain

\begin{eqnarray*}
\langle\Phi_{n}^{\bot}(t)|\hat{F}_{\alpha}(t)\partial_{t}|\Phi_{i}(t)\rangle & = & \sum_{m\neq n}\langle\Phi_{n}^{\bot}(t)|\hat{F}_{\alpha}(t)|\Phi_{m}^{\bot}(t)\rangle\langle\Phi_{m}^{\bot}(t)|\partial_{t}|\Phi_{i}(t)\rangle
+\langle\hat{F}_{\alpha}(t)\rangle_{n}\langle\Phi_{n}^{\bot}(t)|\partial_{t}|\Phi_{i}(t)\rangle.
\end{eqnarray*}
Here, we assume that $c_{\alpha}\neq\langle\hat{F}_{\alpha}\rangle_{n}$
($\forall n$). Taking this term back into Eq.(\ref{eq:b2}), we have

\begin{equation}
\langle\Phi_{n}^{\bot}|\partial_{t}|\Phi_{i}\rangle=\frac{\langle\Phi_{n}^{\bot}|\partial_{t}\hat{F}_{\alpha}|\Phi_{i}\rangle}
{c_{\alpha}-\langle\hat{F}_{\alpha}\rangle_{n}}+\sum_{m\neq n}\frac{\langle\Phi_{n}^{\bot}|\hat{F}_{\alpha}|\Phi_{m}^{\bot}
\rangle}{c_{\alpha}-\langle\hat{F}_{\alpha}\rangle_{n}}
\langle\Phi_{m}^{\bot}|\partial_{t}|\Phi_{i}\rangle.\label{eq:b3}
\end{equation}
For brevity, we have omitted the time arguments on the above equation. By means of same procedure, we obtain
\begin{equation}
\langle\Phi_{m}^{\bot}|\partial_{t}|\Phi_{i}\rangle=\frac{\langle\Phi_{m}^{\bot}|\partial_{t}\hat{F}_{\alpha}
|\Phi_{i}\rangle}{c_{\alpha}-\langle\hat{F}_{\alpha}\rangle_{m}}+\sum_{l\neq m}\frac{\langle\Phi_{m}^{\bot}|\hat{F}_{\alpha}|\Phi_{l}^{\bot}\rangle}{c_{\alpha}-\langle\hat{F}_{\alpha}
\rangle_{m}}\langle\Phi_{l}^{\bot}|\partial_{t}|\Phi_{i}\rangle.\label{eq:b4}
\end{equation}
Let us substitute Eq.(\ref{eq:b4}) into Eq.(\ref{eq:b3}),
\begin{eqnarray*}
\langle\Phi_{n}^{\bot}(t)|\partial_{t}|\Phi_{i}(t)\rangle & = & \frac{\langle\Phi_{n}^{\bot}|\partial_{t}\hat{F}_{\alpha}
|\Phi_{i}\rangle}{c_{\alpha}-\langle\hat{F}_{\alpha}\rangle_{n}}\\
 &  & +\sum_{m\neq n}\frac{\langle\Phi_{n}^{\bot}|\hat{F}_{\alpha}|\Phi_{m}^{\bot}\rangle}{c_{\alpha}-\langle\hat{F}_{\alpha}\rangle_{n}}
 \left(\frac{\langle\Phi_{m}^{\bot}|\partial_{t}\hat{F}_{\alpha}|\Phi_{i}\rangle}{c_{\alpha}-\langle\hat{F}_{\alpha}\rangle_{m}}+\sum_{l\neq m}\frac{\langle\Phi_{m}^{\bot}|\hat{F}_{\alpha}|\Phi_{l}^{\bot}\rangle}{c_{\alpha}-\langle\hat{F}_{\alpha}\rangle_{m}}
 \langle\Phi_{l}^{\bot}|\partial_{t}|\Phi_{i}\rangle\right)\\
 & = & \frac{\langle\Phi_{n}^{\bot}|\partial_{t}\hat{F}_{\alpha}|\Phi_{i}\rangle}{c_{\alpha}-\langle\hat{F}_{\alpha}\rangle_{n}}+\sum_{m\neq n}\frac{\langle\Phi_{n}^{\bot}|\hat{F}_{\alpha}|\Phi_{m}^{\bot}\rangle}{(c_{\alpha}-\langle\hat{F}_{\alpha}\rangle_{n})}\frac{\langle\Phi_{m}^{\bot}
 |\partial_{t}\hat{F}_{\alpha}|\Phi_{i}\rangle}{(c_{\alpha}-\langle\hat{F}_{\alpha}\rangle_{m})}\\
 &  & +\sum_{m\neq n}\sum_{l\neq m}\frac{\langle\Phi_{n}^{\bot}|\hat{F}_{\alpha}|\Phi_{m}^{\bot}\rangle}{c_{\alpha}-\langle\hat{F}_{\alpha}\rangle_{n}}\frac{\langle\Phi_{m}^{\bot}
 |\hat{F}_{\alpha}|\Phi_{l}^{\bot}\rangle}{c_{\alpha}-\langle\hat{F}_{\alpha}\rangle_{m}}\langle\Phi_{l}^{\bot}|\partial_{t}|\Phi_{i}\rangle
\end{eqnarray*}
We can find that $\langle\Phi_{l}^{\bot}|\partial_{t}|\Phi_{i}\rangle$
emerges again. Therefore, we repeat this procedure again and again
until all of bases of the complementary subspace are considered,
\[
\langle\Phi_{n}^{\bot}(t)|\partial_{t}|\Phi_{i}(t)\rangle=\frac{\langle\Phi_{n}^{\bot}|\partial_{t}\hat{F}_{\alpha}|\Phi_{i}\rangle}
{c_{\alpha}-\langle\hat{F}_{\alpha}\rangle_{n}}+\sum_{m\neq n}F_{nm}\frac{\langle\Phi_{m}^{\bot}|\partial_{t}\hat{F}_{\alpha}|\Phi_{i}\rangle}{(c_{\alpha}-\langle\hat{F}_{\alpha}\rangle_{m})}+\sum_{m\neq n}\sum_{l\neq m}F_{nm}F_{ml}\frac{\langle\Phi_{l}^{\bot}|\partial_{t}\hat{F}_{\alpha}|\Phi_{i}\rangle}{(c_{\alpha}-\langle\hat{F}_{\alpha}\rangle_{l})}+...
\]
where
\[
F_{nm}=\frac{\langle\Phi_{n}^{\bot}|\hat{F}_{\alpha}|\Phi_{m}^{\bot}\rangle}{(c_{\alpha}-\langle\hat{F}_{\alpha}\rangle_{n})}.
\]
At this time, the adiabatic condition can be rewrite as
\begin{eqnarray}
\left|\frac{\langle\Phi_{n}^{\bot}(t)|\partial_{t}|\Phi_{i}(t)\rangle}{{(\omega_{ni}+i\Gamma_n)}}\right|&=&\left|\frac{\langle\Phi_{n}
^{\bot}
|\partial_{t}\hat{F}_{\alpha}|\Phi_{i}\rangle}{(\omega_{ni}+i\Gamma_n)(c_{\alpha}-\langle\hat{F}_{\alpha}\rangle_{n})}+\sum_{m\neq n}F_{nm}\frac{\langle\Phi_{m}^{\bot}|\partial_{t}\hat{F}_{\alpha}|\Phi_{i}\rangle}{(\omega_{ni}+i\Gamma_n)(c_{\alpha}-
\langle\hat{F}_{\alpha}\rangle_{m})}\right.\nonumber\\
&&\left.+\sum_{m\neq n}\sum_{l\neq m}F_{nm}F_{ml}\frac{\langle\Phi_{l}^{\bot}|\partial_{t}\hat{F}_{\alpha}|\Phi_{i}\rangle}
{(\omega_{ni}+i\Gamma_n)(c_{\alpha}-\langle\hat{F}_{\alpha}\rangle_{l})}+...\right|
\end{eqnarray}
Thus
\begin{eqnarray*}
\left|\frac{\langle\Phi_{n}^{\bot}(t)|\partial_{t}|\Phi_{i}(t)\rangle}{(\omega_{ni}+i\Gamma_n)}\right| & \leq & \left|\frac{\langle\Phi_{n}^{\bot}|\partial_{t}\hat{F}_{\alpha}|\Phi_{i}\rangle}{(\omega_{ni}+i\Gamma_n)
(c_{\alpha}-\langle\hat{F}_{\alpha}\rangle_{n})}\right|+\sum_{m\neq n}\left|F_{nm}\right|\left|\frac{\langle\Phi_{m}^{\bot}|\partial_{t}\hat{F}_{\alpha}|\Phi_{i}\rangle}
{(\omega_{ni}+i\Gamma_n)(c_{\alpha}-\langle\hat{F}_{\alpha}\rangle_{m})}\right|\nonumber\\
&&+\sum_{m\neq n}\sum_{l\neq m}\left|F_{nm}\right|\left|F_{ml}\right|\left|\frac{\langle\Phi_{l}^{\bot}|\partial_{t}\hat{F}_{\alpha}
|\Phi_{i}\rangle}{(\omega_{ni}+i\Gamma_n)(c_{\alpha}-\langle\hat{F}_{\alpha}\rangle_{l})}\right|+...
\end{eqnarray*}
Here, we have used the facts that any complex number $a$
and $b$ fulfill $|ab|\leq|a||b|$ and $|a+b|\leq|a|+|b|$. Here,
we assume that all of $\{|F_{nm}|,\forall m,n\}$ are bounded by a finite real number
$F_{\text{Max}}$. Replacing all $\left|\frac{\langle\Phi_{n}^{\bot}|\partial_{t}\hat{F}_{\alpha}|\Phi_{i}\rangle}
{\omega_{ni}(c_{\alpha}-\langle\hat{F}_{\alpha}\rangle_{n})}\right|$ terms in above equation by the
maximum among them, we can rewrite the adiabatic condition Eq.(\ref{eq:maxnj}) into
\begin{eqnarray}
\text{Max}_{\{n,i\}}\left|\langle\Phi_{n}^{\bot}(t)|\partial_{t}|\Phi_{i}(t)\rangle\right| & \leq & \frac{1}{N-M}\text{Max}_{\{n,i\}}\left|\frac{\langle\Phi_{n}^{\bot}|\partial_{t}\hat{F}_{\alpha}
|\Phi_{i}\rangle}{(\omega_{ni}+i\Gamma_n)(c_{\alpha}-\langle\hat{F}_{\alpha}\rangle_{n})}\right|\nonumber\\
&\times&[N-M+(N-M)(N-M-1)F_{\text{Max}}\nonumber\\
&& +(N-M)(N-M-1)(N-M-2)F_{\text{Max}}^{2}\nonumber\\
&&+...+(N-M)!F_{\text{Max}}^{M-1}]\\
 & = & \left(\frac{\sum_{a=0}^{N-M-1}P_{N-M}^{a+1}(F_{\text{Max}})^{a}}
 {N-M}\right)\text{Max}_{\{n,i\}}\left|\frac{\langle\Phi_{n}^{\bot}|\partial_{t}\hat{F}_{\alpha}|\Phi_{i}\rangle}
 {(\omega_{ni}+i\Gamma_n)(c_{\alpha}-\langle\hat{F}_{\alpha}\rangle_{n})}\right|,\label{eq:maxf}
\end{eqnarray}
where $P_{N-M}^{a+1}$ is the number of permutation.

\section{ The lower bound of the Purity} \label{appendix c}

We start with the inequality mentioned in Eq.(\ref{eq:0-p}), i.e.,
\begin{eqnarray*}
p(t)-1&\geq&4\text{Re}\left\{\sum_{ijm}-i\bar{\rho}_{ji}\bar{\rho}_{im}\frac{\langle\Phi_{m}^{\bot}
|\partial_{t}|\Phi_{j}\rangle}{\omega_{mj}+i\Gamma_{m}}\exp\left(i\int_{0}^{t}(\omega_{mj}+i\Gamma_{m})
d\tau\right)\right.\\&&+i\int_{0}^{t}\partial_{t'}\bar{\rho}_{ji}\bar{\rho}_{im}\frac{\langle\Phi_{m}^{\bot}|
\partial_{t'}|\Phi_{j}\rangle}{\omega_{mj}+i\Gamma_{m}}\exp\left(i\int_{0}^{t'}(\omega_{mj}+i\Gamma_{m})
d\tau\right)dt'\\&&+i\int_{0}^{t}\bar{\rho}_{ji}\partial_{t'}\bar{\rho}_{im}\frac{\langle\Phi_{m}^{\bot}|
\partial_{t'}|\Phi_{j}\rangle}{\omega_{mj}+i\Gamma_{m}}\exp\left(i\int_{0}^{t'}(\omega_{mj}+i\Gamma_{m})
d\tau\right)dt'\\&&\left.+i\int_{0}^{t}\bar{\rho}_{ji}\bar{\rho}_{im}\partial_{t'}\left(\frac{\langle\Phi_{m}^{\bot}|
\partial_{t'}|\Phi_{j}\rangle}{\omega_{mj}+i\Gamma_{m}}\right)\exp\left(i\int_{0}^{t'}(\omega_{mj}+i\Gamma_{m})d\tau\right)dt'\right\}.
\end{eqnarray*}
Since the lower bound considered here is under adiabatic condition, we
check the dynamical equation if the adiabatic approximation are reached.
After transforming by the operator Eq.(\ref{eq:U}), the master
equation can be rewritten as Eq.(\ref{eq:liou3}), i.e.,
\[
\partial_{t}\bar{\rho}(t)=-i[\bar{H}_{\text{eff}}+\bar{G},\bar{\rho}]+\sum_{\alpha}
\left(\tilde{F}_{\alpha}\bar{\rho}\tilde{F}_{\alpha}^{\dagger}-\frac{1}{2}
\{\tilde{F}_{\alpha}^{\dagger}\tilde{F}_{\alpha},\bar{\rho}\}\right).
\]
Hence, we can obtain

\begin{eqnarray*}
\partial_{t}\bar{\rho}_{ji}&=&\langle\Phi_{j}(0)|\partial_{t}\bar{\rho}(t)|\Phi_{i}(0)\rangle\\
&=&-i\langle\Phi_{j}(0)|[\bar{H}_{\text{eff}}+\bar{G},\bar{\rho}]|\Phi_{i}(0)\rangle+\sum_{\alpha}
\langle\Phi_{j}(0)|\left(\tilde{F}_{\alpha}\bar{\rho}\tilde{F}_{\alpha}^{\dagger}
-\frac{1}{2}\{\tilde{F}_{\alpha}^{\dagger}\tilde{F}_{\alpha},\bar{\rho}\}\right)|\Phi_{i}(0)\rangle\\
&=&-i\langle\Phi_{j}(0)|((\bar{H}_{\text{eff}}+\bar{G})\bar{\rho}-\bar{\rho}
(\bar{H}_{\text{eff}}+\bar{G}))]|\Phi_{i}(0)\rangle\\
&=&\sum_{k}\left(\exp\left(i\int_{0}^{t}\omega_{jk}d\tau\right)\langle\partial_{t}\Phi_{j}(t)|\Phi_{k}(t)\rangle\bar{\rho}_{ki}
-\exp\left(i\int_{0}^{t}\omega_{ki}\tau\right)\bar{\rho}_{jk}\langle\partial_{t}\Phi_{k}(t)
|\Phi_{i}(t)\rangle\right)\\
&&+\sum_{n}\left(\exp\left(i\int_{0}^{t}\omega_{jn}-i\langle\hat{\Gamma}(\tau)
\rangle_{n}d\tau\right)\langle\partial_{t}\Phi_{j}(t)|\Phi_{n}^{\bot}(t)\rangle\bar{\rho}_{ni}\right.\\
&&\left.-\exp\left(i\int_{0}^{t}(\omega_{ni}+i\langle\hat{\Gamma}(\tau)\rangle_{n})d\tau\right)\bar{\rho}_{jn}
\langle\partial_{t}\Phi_{n}^{\bot}(t)|\Phi_{i}(t)\rangle\right).
\end{eqnarray*}
and
\begin{eqnarray*}
\partial_{t}\bar{\rho}_{im}&=&\langle\Phi_{i}(0)|\partial_{t}\bar{\rho}(t)|\Phi_{m}^{\bot}(0)\rangle\\
&=&\langle\Phi_{i}(0)|\left(-i[\bar{H}_{\text{eff}}+\bar{G},\bar{\rho}]+\sum_{\alpha}\left(\tilde{F}_{\alpha}
\bar{\rho}\tilde{F}_{\alpha}^{\dagger}-\frac{1}{2}\{\tilde{F}_{\alpha}^{\dagger}\tilde{F}_{\alpha},\bar{\rho}\}\right)\right)
|\Phi_{m}^{\bot}(0)\rangle\\
&=&-i\left(\sum_{j}\exp\left(i\int_{0}^{t}\langle\hat{H}_{\text{eff}}^{0}(\tau)\rangle_{i}-\langle\hat{H}_{\text{eff}}^{0}(\tau)
\rangle_{j}d\tau\right)\langle\partial_{t}\Phi_{i}(t)|\Phi_{j}(t)\rangle\bar{\rho}_{jm}\right.\\
&&+\sum_{n}\exp\left(i\int_{0}^{t}
\omega_{in}-i\langle\hat{\Gamma}(\tau)\rangle_{n}d\tau\right)\langle\partial_{t}\Phi_{i}(t)|\Phi_{n}^{\bot}(t)\rangle
\bar{\rho}_{nm}\\
&&-\sum_{j}\exp\left(i\int_{0}^{t}\omega_{jm}-i\langle\hat{\Gamma}(\tau)\rangle_{m}d\tau\right)\bar{\rho}_{ij}
\langle\partial_{t}\Phi_{j}(t)|\Phi_{m}^{\bot}(t)\rangle\\
&&\left.-\sum_{n}\exp\left(i\int_{0}^{t}(\omega_{nm}+i\langle\hat{\Gamma}
(\tau)\rangle_{n}-i\langle\hat{\Gamma}(\tau)\rangle_{m})d\tau\right)\bar{\rho}_{in}\langle\partial_{t}\Phi_{n}^{\bot}(t)
|\Phi_{m}^{\bot}(t)\rangle\right)\\
&&-\sum_{n\neq m}\sum_{n}\bar{\rho}_{in}\langle\Phi_{n}^{\bot}(t)|\hat{\Gamma}(t)|\Phi_{m}^{\bot}(t)\rangle\exp\left(i\int_{o}^{t}\omega_{nm}
+i\langle\hat\Gamma(\tau)\rangle_{n}-i\langle\hat\Gamma(\tau)\rangle_{m}d\tau\right).
\end{eqnarray*}
In this expression, we have also omitted the "jump" terms $\tilde{F}_{\alpha}\bar{\rho}\tilde{F}_{\alpha}^{\dagger}$ with the same reason in Appendix.\ref{sec:Appendix-A}. Taking them into Eq.(\ref{eq:0-p}), the lower bound can be rewritten
as
\begin{eqnarray*}
p(t)-1 & \geq &4\text{Re}\left\{\sum_{ijm}-i\rho_{ji}\rho_{im}\frac{\langle\Phi_{m}^{\bot}(t)|\partial_{t}|\Phi_{j}(t)\rangle}
{\omega_{mj}+i\Gamma_{m}}\right.\\
&&+\int_{0}^{t}\sum_{k}\langle\partial_{t}\Phi_{j}(t)|\Phi_{k}(t)\rangle\rho_{ki}\rho_{im}
\frac{\langle\Phi_{m}^{\bot}(t)|\partial_{t'}|\Phi_{j}(t)\rangle}{\omega_{mj}+i\Gamma_{m}}dt'\\
&&+\int_{0}^{t}\sum_{n}
\langle\partial_{t}\Phi_{j}(t)|\Phi_{n}^{\bot}(t)\rangle\rho_{ni}\rho_{im}\frac{\langle\Phi_{m}^{\bot}(t)|\partial_{t'}
|\Phi_{j}(t)\rangle}{\omega_{mj}+i\Gamma_{m}}dt'\\
&&-\int_{0}^{t}\sum_{n}\rho_{jn}\langle\partial_{t}\Phi_{n}^{\bot}(t)
|\Phi_{i}(t)\rangle\rho_{im}\frac{\langle\Phi_{m}^{\bot}(t)|\partial_{t'}|\Phi_{j}(t)\rangle}{\omega_{mj}+i\Gamma_{m}}dt'\\
&&+\int_{0}^{t}\sum_{n}\rho_{ji}\langle\partial_{t}\Phi_{i}(t)|\Phi_{n}^{\bot}(t)\rangle\rho_{nm}\frac{\langle\Phi_{m}^{\bot}(t)|
\partial_{t'}|\Phi_{j}(t)\rangle}{\omega_{mj}+i\Gamma_{m}}dt'\\&&-\int_{0}^{t}\sum_{k}\rho_{ji}\rho_{ik}\langle\partial_{t}
\Phi_{k}(t)|\Phi_{m}^{\bot}(t)\rangle\frac{\langle\Phi_{m}^{\bot}(t)|\partial_{t'}|\Phi_{j}(t)\rangle}{\omega_{mj}+i\Gamma_{m}}dt'\\
&&-\int_{0}^{t}\sum_{n}\rho_{ji}\rho_{in}\langle\partial_{t}\Phi_{n}^{\bot}(t)|\Phi_{m}^{\bot}(t)\rangle\frac{\langle
\Phi_{m}^{\bot}(t)|\partial_{t'}|\Phi_{j}(t)\rangle}{\omega_{mj}+i\Gamma_{m}}dt'\\&&-\int_{0}^{t}\sum_{n\neq m}\rho_{ji}\rho_{in}\langle\Phi_{n}^{\bot}|\bar{\Gamma}|\Phi_{m}^{\bot}\rangle\frac{\langle\Phi_{m}^{\bot}(t)|
\partial_{t'}|\Phi_{j}(t)\rangle}{\omega_{mj}+i\Gamma_{m}}dt'\\&&\left.+i\int_{0}^{t}\rho_{ji}\rho_{im}\frac{\partial}{\partial t'}
\left(\frac{\langle\Phi_{m}^{\bot}(t)|\partial_{t'}|\Phi_{j}(t)\rangle}{\omega_{mj}+i\Gamma_{m}}\right)dt'\right\},
\end{eqnarray*}
with $\rho_{ij}=\langle \Phi_i (t)|\hat \rho (t)|\Phi_j(t)\rangle$ and $\rho_{in}=\langle \Phi_i (t)|\hat \rho (t)|\Phi_n^\bot(t)\rangle$. In the general case, although it is difficult to estimate exactly
the values of the integrals in above equation, it is still possible
to obtain bounds on the integrals, which will lead to the lower bound.
Noting that $|{\rho}_{ij}|\leq1$ and $|\exp(i\int_{0}^{t'}\omega_{nj}d\tau)|=1$,
we have
\begin{eqnarray*}
1-p(t)&\leq&4\sum_{jm}(M\left|\frac{\langle\Phi_{m}^{\bot}|\partial_{t}|\Phi_{j}\rangle}{\omega_{mj}+i\Gamma_{m}}\right|\\
&&+M\int_{0}^{t}\left(\sum_{k}\left|\langle\partial_{t}\Phi_{j}(t)|\Phi_{k}(t)\rangle\right|+\sum_{n}\left|\langle\partial_{t}
\Phi_{j}(t)|\Phi_{n}^{\bot}(t)\rangle\right|\right)\left|\frac{\langle\Phi_{m}^{\bot}|\partial_{t'}|\Phi_{j}\rangle}
{\omega_{mj}+i\Gamma_{m}}\right|dt'\\
&&+\sum_{i}\sum_{n}\int_{0}^{t}\left(|\langle\partial_{t}\Phi_{n}^{\bot}(t)|\Phi_{i}(t)\rangle|+
|\langle\partial_{t}\Phi_{i}(t)|\Phi_{n}^{\bot}(t)\rangle|\right)\left|\frac{\langle\Phi_{m}^{\bot}|\partial_{t'}
|\Phi_{j}\rangle}{\omega_{mj}+i\Gamma_{m}}\right|dt'\\
&&+M\int_{0}^{t}\left(\sum_{k}|\langle\partial_{t}\Phi_{k}(t)
|\Phi_{m}^{\bot}(t)\rangle+\sum_{n}|\langle\partial_{t}\Phi_{n}^{\bot}(t)|\Phi_{m}^{\bot}(t)\rangle+\sum_{n\neq m}|\langle\Phi_{n}^{\bot}|\bar{\Gamma}|\Phi_{m}^{\bot}\rangle|\right)\left|\frac{\langle\Phi_{m}^{\bot}|\partial_{t'}
|\Phi_{j}\rangle}{\omega_{mj}+i\Gamma_{m}}\right|dt'\\
&&+M\int_{0}^{t}\left|\partial_{t'}\left(\frac{\langle\Phi_{m}^{\bot}
|\partial_{t'}|\Phi_{j}\rangle}{\omega_{mj}+i\Gamma_{m}}\right)\right|dt')\\
&=&4M\sum_{jm}(\left|\frac{\langle\Phi_{m}^{\bot}
|\partial_{t}|\Phi_{j}\rangle}{\omega_{mj}+i\Gamma_{m}}\right|+\int_{0}^{t}(A_{j}+B_{m}+C)\left|\frac{\langle\Phi_{m}^{\bot}
|\partial_{t'}|\Phi_{j}\rangle}{\omega_{mj}+i\Gamma_{m}}\left|dt'+\int_{0}^{t}\right|\frac{\partial}{\partial t'}\left(\frac{\langle\Phi_{m}^{\bot}|\partial_{t'}|\Phi_{j}\rangle}{\omega_{mj}+i\Gamma_{m}}\right)\right|dt')
\end{eqnarray*}
where
\begin{eqnarray*}
A_{j}&=&\sum_{k}|\langle\partial_{t}\Phi_{j}(t)|\Phi_{k}(t)\rangle|+\sum_{n}|\langle\partial_{t}\Phi_{j}(t)
|\Phi_{n}^{\bot}(t)\rangle|\\
B_{m}&=&\sum_{k}|\langle\partial_{t}\Phi_{k}(t)|\Phi_{m}^{\bot}(t)\rangle+\sum_{n}|\langle\partial_{t}
\Phi_{n}^{\bot}(t)|\Phi_{m}^{\bot}(t)\rangle+\sum_{n\neq m}|\langle\Phi_{n}^{\bot}|\bar{\Gamma}|\Phi_{m}^{\bot}\rangle|\\
C&=&\sum_{i}\sum_{n}\left(|\langle\partial_{t}\Phi_{n}^{\bot}(t)|\Phi_{i}(t)\rangle|+
|\langle\partial_{t}\Phi_{i}(t)|\Phi_{n}^{\bot}(t)\rangle|\right)/M
\end{eqnarray*}
In the derivation, we have used the fact that $|ab|\leq|a||b|$ and
$|a+b|\leq|a|+|b|$ with arbitrary complex numbers $a$ and $b$.
By uniting like terms, we reach Eq.(\ref{eq:2-p}) in Sec.\ref{sub:The-Lower-Bound}.\end{widetext}

\end{document}